\documentclass{article}

\usepackage{arxiv}

\usepackage[utf8]{inputenc} 
\usepackage[T1]{fontenc}    
\usepackage{hyperref}       
\usepackage{url}            
\usepackage{booktabs}       
\usepackage{amsfonts}       
\usepackage{nicefrac}       
\usepackage{microtype}      
\usepackage{subcaption}
\usepackage{amsmath}
\usepackage{rotating}
\usepackage{cleveref}       
\usepackage{graphicx}
\usepackage[numbers]{natbib}
\usepackage{doi}

\title{Opinion Dynamics in Two-Layer Networks with Hypocrisy}

\date{March 11, 2023}

\author{ \href{https://orcid.org/0000-0002-1166-7578}{\includegraphics[scale=0.06]{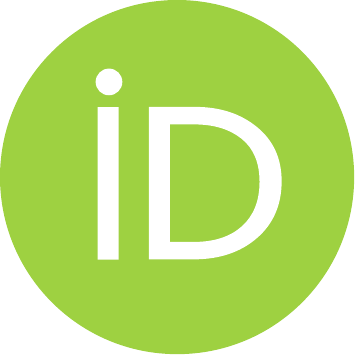}\hspace{1mm}Chi Zhao} \\ 
	Saint Petersburg State University,\\
	7/9 Universitetskaya nab.,\\
	Saint Petersburg, 199034, Russia\\
	\texttt{st081292@student.spbu.ru} \\
	\And
	\href{https://orcid.org/0000-0003-3976-7180}{\includegraphics[scale=0.06]{orcid.pdf}\hspace{1mm}Elena Parilina} \\
	Saint Petersburg State University,\\
	7/9 Universitetskaya nab.,\\
	Saint Petersburg, 199034, Russia\\
	School of Mathematics and Statistics,\\
	Qingdao University,\\
	Qingdao, 266071, PR China\\
	\texttt{e.parilina@spbu.ru} \\
}

\renewcommand{\shorttitle}{\textit{Opinion Dynamics in Two-Layer Networks with Hypocrisy}}

\hypersetup{
pdftitle={Opinion Dynamics in Two-Layer Networks with Hypocrisy},
pdfsubject={Opinion Dynamics},
pdfauthor={Chi Zhao, Elena Parilina},
pdfkeywords={Opinion Dynamics, Voter Model, Concealed Voter Model, General Concealed Voter Model},
}

\begin{document}
\maketitle

\begin{abstract}
We propose a general concealed voter model (GCVM), in which individuals interact in two layers and can exchange their opinions in the internal layer. This interaction is not allowed in a concealed voter model (CVM). By exchanging opinions in the internal layer we mean that individuals share their real or internal opinions with their close friends. The process of opinion formation in GCVM is presented in the paper. We make the series of numerical simulations of GCVM with different network structures (both external and internal) and get some counterintuitive conclusions. For instance, we find out that sometimes with a relatively simple network structure of an external layer the consensus within the individuals' opinions cannot be reached, and if individuals in the network are not good at expressing their opinions publicly (in an external layer), exchanging opinions with their close friends (in an internal layer) is almost useless.
\end{abstract}


\section{Introduction}\label{introduction}

Modeling the opinion dynamics is a research area mainly focusing on studying the evolution of opinions in the social networks caused by individuals' interactions.
Opinion dynamics models can currently be divided into two main groups: macroscopic and microscopic models. Macroscopic models  examine social networks using statistical-physical methods and applying probability and statistics theories to analyze how the distribution of opinion evolves, e.g., the Ising model \cite{mckeehan1925contribution} and voter model \cite{holley1975ergodic}. The Ising model has a long history in statistical physics \cite{noorazar2020recent}. The Sznajd model \cite{sznajd2000opinion} is one of the well-known modification of the Ising model. In each round described in the Sznajd model, a pair of agents $a_i$ and $a_{i+1}$ is selected for interaction to influence the nearest neighbors, i.e. the agents $a_{i-1}$ and $a_{i+2}$. In a voter model, a random agent $a_i$ is chosen, who chooses a random neighbor, and this neighbor adopts $a_i$'s opinion.

Microscopic models directly describe how individuals' opinions evolve from social individuals' perspectives, e.g., see the DeGroot model \cite{degroot1974reaching}, the Friedkin-Johnsen (F-J) model \cite{friedkin1990social}, and bounded confidence models \cite{rainer2002opinion, deffuant2000mixing}.
In the DeGroot model, each individual updates his opinion based on his own and neighbors' opinions. The F-J model is one of the major extension of the DeGroot model, and in the F-J model, the presence of stubborn-agents extends the DeGroot model. In the F-J model, actors can also factor their initial prejudices into every iteration of opinion \cite{parsegov2016novel}. The possibility to control the agents' opinions by non-members of network is considered in \cite{sedakovrogov, mazalovparilina}. A bounded confidence model (BCM) is a model, in which agents ignore the opinions that are very far from their own opinions \cite{noorazar2020recent}. The BCM includes two essential models: the Deffuant-Weisbuch model (D-W) proposed in paper \cite{deffuant2000mixing}, and the Hegselman-Krause (H-K) model introduced in the work \cite{rainer2002opinion}. In the D-W model, two individuals $a_i, a_j$ are randomly chosen, and they  determine whether to interact according to the bounded confidence \cite{zha2020opinion}. The H-K model is also an extension of the DeGroot model, in which it is assumed that every individual in the network has a confidence bound, and the individual's opinion can be affected only by others within their own confidence bound. Therefore, the opinion updating rule of the H-K model is state-related. The Krasnoshchekov's model (K) of opinion dynamics in the society represented by one layer is introduced in \cite{Krasno} and then it is examined in \cite{Kozitsin}. This opinion dynamics can be reduced to the F-J dynamics. In fact, if the opinion updating rule presented in K model is applied, the corresponding dynamics gives the converge of the agents' opinions to some terminal opinions. Therefore, a consensus is reached.

The presence of a group of agents who can manipulate the opinions of the society is examined in \cite{Roland}. The agents are assumed to be heterogeneous in this model, taking into account that the group of the so-called leaders knows the initial opinions of all the agents, while the rest of the agents does not know this information.

According to \cite{dong2018survey}, opinion dynamics models are usually composed of a few essential elements: (i)
opinion expression formats defining how to represent opinion mathematically, (ii) fusion rule determining how individuals interact with each other, and (iii) opinion dynamics environments, that is the structure of such a social network.

In a social network, individuals neither fully accept nor completely ignore the opinions of other individuals.
To a certain extent, they consider these opinions in forming their new opinion in a process defined by a fusion rule.
Through group interaction, individuals continuously update and integrate their opinions on the same issue. Eventually, there are three varieties of stabilized fusion results: consensus, polarization, fragmentation, and one unstable fusion result, that is oscillation \cite{rainer2002opinion}.

The basic voter model proposed by Richard and Thomas is called  BVM \cite{holley1975ergodic}, the concealed voter model proposed by Gastner et al. as CVM \cite{gastner2018consensus, gastner2019impact}.
The structure of BVM is based on a complete network, and the general assumption of this model is that individuals always express their opinions publicly. Therefore, the fusion rule of BVM is quite intuitive, that is, selecting an individual and his neighbor randomly, and the individual adopts his neighbor's opinion. CVM assumes that the social network is divided into external and internal layers, and the individuals feel free to conceal or publicly express their opinions. The external layer of CVM is a complete network, and each node is the external layer linked with a node in the internal layer. Moreover, there are no connections between node in the internal layer. Therefore, internal interaction is not allowed in CVM.
In this paper, we assume that the individuals can interact in the internal layer, which is an assumption making our model different from CVM. We propose this idea motivated by the fact that the individuals always share their real opinions with their close friends.  The multilevel approach to model the structure of the society including team networks in companies is widely spread in the models of industrial organization \cite{Donati}. Therefore, we also incorporate this multilevel communication approach in the model of opinion dynamics, and call CVM with possible interactions in the internal layer as GCVM.

The rest of this paper is organized as follows. Section 2 introduces BVM and CVM. In Section 3, we present GCVM with different network structures. The Monte Carlo method is used to simulate BVM, CVM and GCVM with different network structures in Section 4, where we also provide our observations based on the simulations. Finally, we briefly conclude our current work and outlook for the future research.

\section{Basic and Concealed Voter Models}\label{sec2}

We assume that the society is represented by the agents or individuals who have opinions on the topic. The opinion is binary, i.e., each individual's opinion is red or blue. We introduce the models of opinion dynamics taking into account the network structure the individuals form within the social network which we call general concealed voter model (GCVM). Before introducing GCVM, we briefly introduce a basic voter model (BVM) and concealed voter models (CVM) (see \cite{gastner2018consensus,gastner2019impact}).

We call a classical voter model as a basic voter model. 
BVM assumes that everyone in a network can express his opinion publicly, so there is only one layer of information exchange and a unique opinion of an individual about the topic in this model of opinion dynamics. In CVM described in Section \ref{CVM}, individuals can either express or conceal their real (hidden) opinion publicly. An individual has  both publicly expressed and private opinions about the topic, and the private opinion is unknown to other individuals. If an individual conceals his real opinion, i.e. his publicly known opinion is different from the private one, we call him \textit{hypocrisy}: an individual with cognitive dissonance \cite{festinger1962theory}. There are two layers (external and internal) in the CVM but the information exchange in the internal layer is not allowed.  

We focus on the time when groups under different network structures reach consensus, this time is called consensus time, and the winning rate of an opinion after a series of simulations.

\subsection{Basic Voter Model (BVM)}\label{BVM}

Suppose we have a predefined network $G$ representing the communication or network structure connecting players in the society. Denote the number of individuals in  network $G$ by $N$. We examine the evolution of network in continuous time. We use the following notations:
\begin{itemize}
  \item $\omega_{ext}(\alpha,t) \in \{0,1\} $ is the opinion of individual $\alpha$ at time $t$
    (0 is represented by blue and 1 is represented by red color);
  \item $c$ is a copy rate, that is the probability of individuals adapt his neighbor's opinion;
  \item $r$ is the number of individuals with red opinion;
  \item $\rho$ is the proportion of individuals with red opinion in the population
    (the strength of red opinion), that is $\rho=r/N$.
  \item $T_{cons}$ is the consensus time, in BVM it means the time required for all individuals to form the same opinion ($\rho=1$ or $\rho=0$).
\end{itemize}

We use the proportion of red opinion to represent the state of such a BVM system, and $\rho\in\{0,\frac{1}{N},\ldots,\frac{N-1}{N},1\}$. It is assumed that any individual can change his opinion as a result of a stochastic event realization.
We suppose that the interval of such an event obeys the exponential distribution with arrival rate $\lambda$.
Then we define the following stochastic matrix with elements $P(\rho,\cdot)$ for each $\rho \notin \{0,1\}$, where $P(\rho,\rho^{\prime})$ is the probability of transition of BVM system from state $\rho$ to state $\rho^{\prime}$ after the event happens. If $\rho \in \{0,1\}$, we have $P(0,0)=1$ and $P(1,1)=1$ meaning that if all individuals in the network have the same opinion, they will never change it. 
The proportion of red opinion $\rho$ can be changed to $\rho\pm\frac{1}{N}$ or remains the same with positive probabilities:
\begin{gather}
  \begin{aligned}
    \label{eq:BVMstochasticMatrix}
    & P\big(\rho,\rho+\frac{1}{N}\big) = \frac{c(N-r)r}{N(N-1)}=\frac{c(1-\rho)r}{N-1},\\
    & P\big(\rho,\rho-\frac{1}{N}\big) = \frac{cr(N-r)}{N(N-1)}=\frac{cr(1-\rho)}{N-1},\\
    & P(\rho,\rho) = 1-2\frac{cr(1-\rho)}{N-1}.
  \end{aligned}
\end{gather}

If we consider the change of the BVM system state as a random event,
then the change rate per unit time in this system can be represented as
the sum of positive direction change rate $\lambda_{+}$ and negative direction change rate $\lambda_{-}$, where
\begin{equation}
  \begin{aligned}
    \label{eq:BVMchangeRate}
    &\lambda_{-}=cr\frac{N-r}{N}=\lambda_{+}=c(N-r)\frac{r}{N},\\
    &\lambda= \lambda_{+}+\lambda_{-}=2c(N-r)\frac{r}{N}.
  \end{aligned}
\end{equation}

Therefore, we have a system with an arrival rate $\lambda$ per unit time. The following steps show how individuals interact with each other in BVM:

\begin{enumerate}
  \item \textbf{Initialization:} Given fraction $\rho$ of the individuals with red opinion, other individuals hold blue opinion. Initialize the time $t\rightarrow 0$.
  \item \textbf{Iteration:}
    \begin{enumerate}
                \item[a.] Choose a ``focal'' individual $f$ uniformly at random from all of $N$ individuals.
                \item[b.] Pick a neighbor $a$ of the focal individual uniformly at random from all of his neighbors.
                \item[c.] Generate a temporary varialbe $\lambda_{t}=\lambda \cdot u$, where $u \sim U(0,1)$, individual $f$ adopts neighbor $a$'s opinion according to $\lambda_{t}$ by the following way:
                  \begin{enumerate}
                    \item[i.] The state of this system will be changed from $\rho$ to $\rho+\frac{1}{N}$, if $\lambda_{t}<\lambda_{+}$;
                    \item[ii.] The state of this system will be changed from $\rho$ to $\rho-\frac{1}{N}$, if $\lambda_{+} \leq \lambda_{t} \leq\lambda_{+}+\lambda_{-}=\lambda$,
                  \end{enumerate}
                \item[d.]  We increase the time by a random number $ \Delta t $ drawn from an exponential distribution with mean $ \frac{1}{\lambda} $, $ t \rightarrow t+\Delta t $.
                \item[e.]  If the group has reached a consensus, we set $t  \rightarrow T_{\text {cons}}^{(\text {BVM})}$ and terminate. Otherwise we go back to step a.
    \end{enumerate}
\end{enumerate}

\begin{figure}[htpb]
  \centering
  \includegraphics[width=0.6\textwidth]{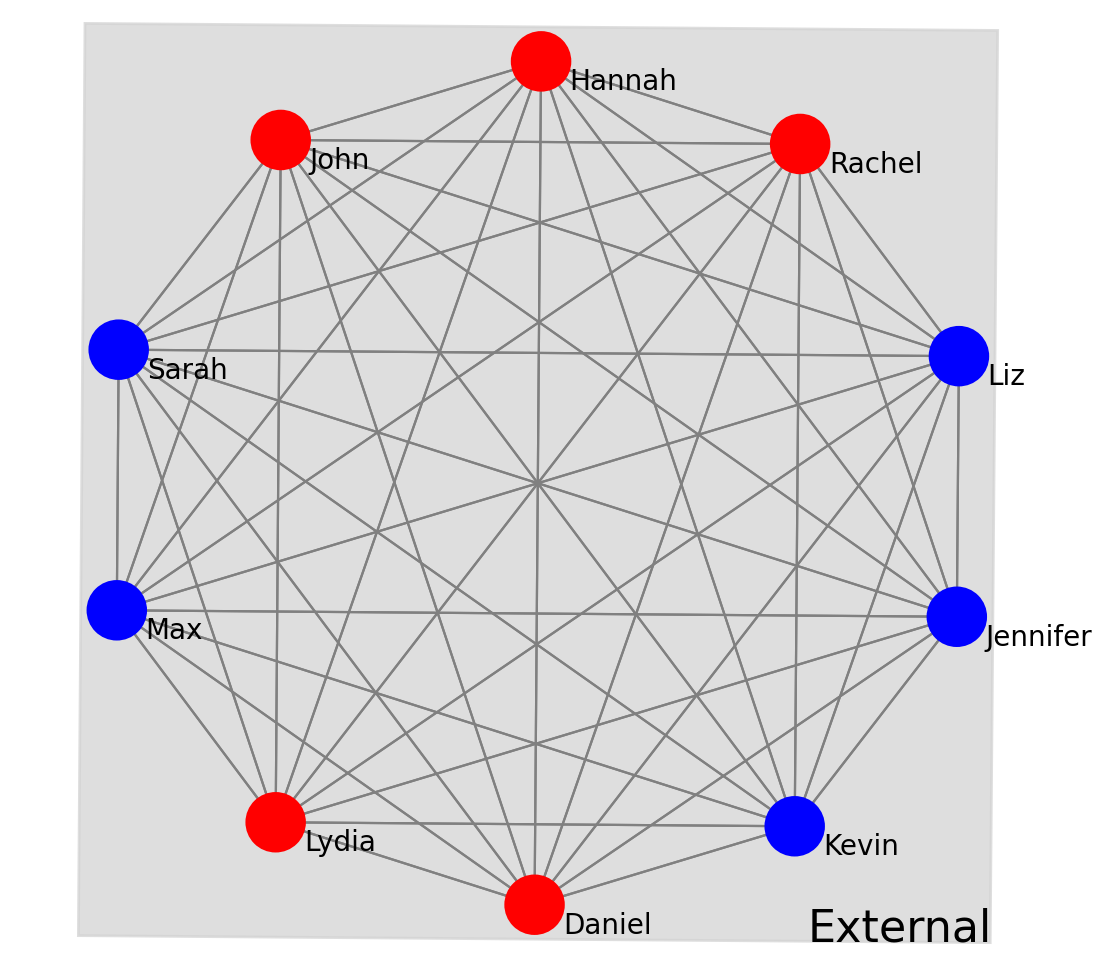}
  \caption{A complete network connecting 10 individuals in BVM}
  \label{fig:BVM}
\end{figure}

Fig.~\ref{fig:BVM} represents the network of BVM with $10$ individuals on the  complete network.

\subsection{Concealed Voter Model (CVM)}\label{CVM}

\begin{figure}[htpb]
	\centering
	\includegraphics[width=0.7\textwidth]{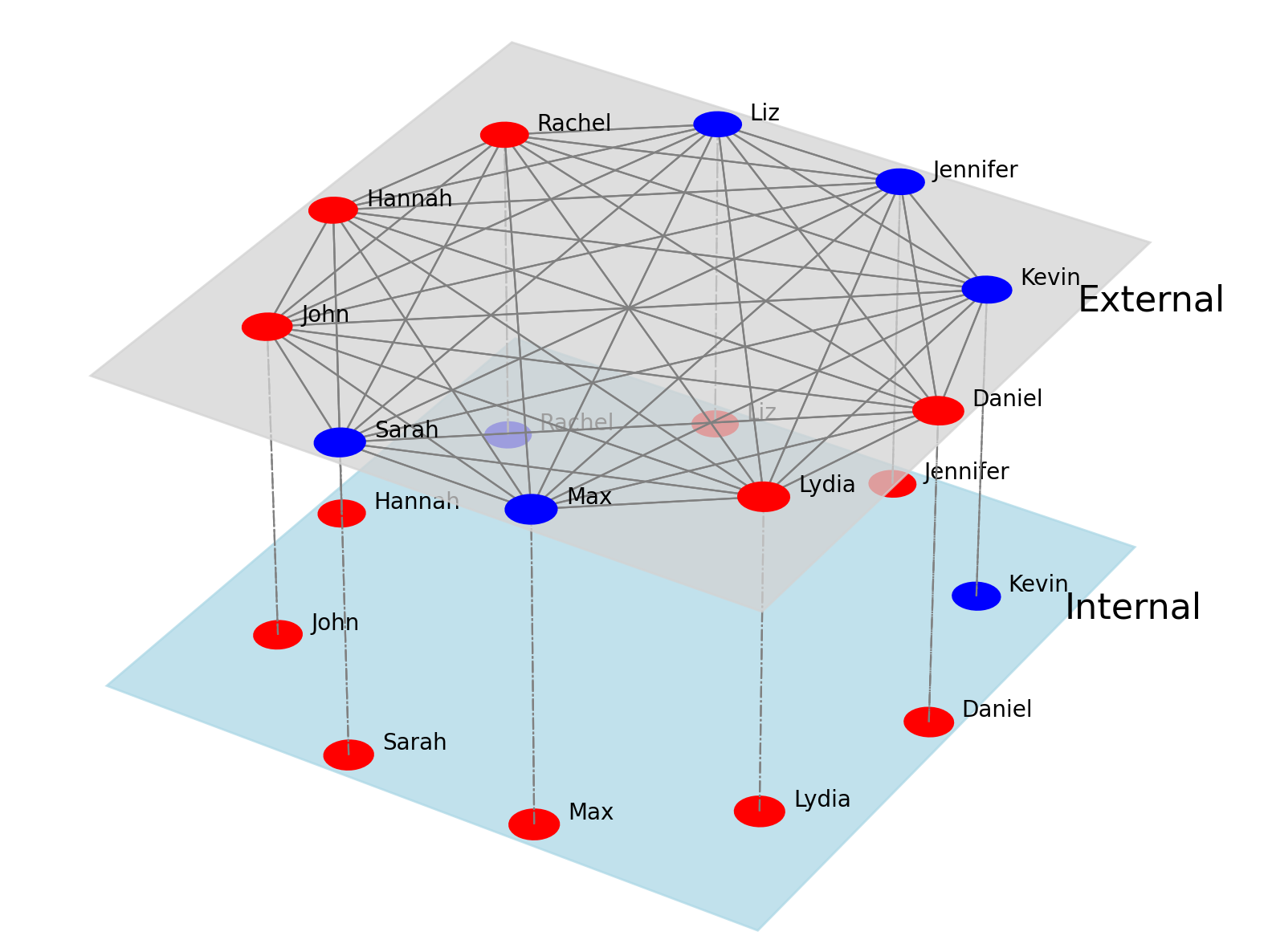}
	\caption{A two-layer network structure in CVM with 10 individuals}
	\label{fig:CVM}
\end{figure}

In CVM the predefined networks $G_1$ and $G_2$ are given with the same set of $N$ individuals. Individual $i$ is represented by node $E_i$ in network $G_1$ called an external layer, and by node $I_i$ in network $G_2$ called an internal layer. Therefore, each individual is represented by a pair of nodes $(E_i,I_i)$, $i=1,\ldots,N$. We should notice that there is no restriction on the network structure $G_1$ but network $G_2$ is always an empty network with $N$ individuals. The two-layer network structure satisfies the following properties:
\begin{enumerate}
	\item There are no links between nodes $I_i$ and $I_j$, $\forall i,j=1,\ldots,N$.
	\item There must exist a link between $E_i$ and $I_i$, $\forall i=1,\ldots,N$.
\end{enumerate}
Fig. \ref{fig:CVM} represents such a two-layer network with 10 individuals when the external layer is a complete network.

In CVM, we use $R, B$ ($r, b$)  to represent individuals' external (internal)  red and blue opinions respectively. Denote the state set of an individual's opinion by $S=\{Rr,Rb,Bb,Br\}$. We use the following notations for CVM systems:
\begin{itemize}
  \item $\omega(\alpha, t)\in S$ is the state of individual $\alpha$'s opinion at time $t$;
  \item $c$ is a copy rate (equivalent to the one in the BVM);
  \item $e$ is the externalization rate, that is the probability of hypocrisy choosing to publicly express his internal opinion;
  \item $i$ is the internalization rate, that is the probability of hypocrisy choosing to accept his external opinion;
  \item $r_e$ is the number of individuals with external red opinion;
  \item $r_i$ is the number of individuals with internal red opinion;
  \item $r$ is the number of individuals with both external and internal red opinion;
  \item $\rho_{r_e}, \rho_{r_i}, \rho_{r}$ -- the rate of external red, internal red and both external and internal red opinions.
  \item $T_{cons}$ is the consensus time, in (G)CVM it means the time required for all individuals to form the same opinion in internal and external layers (i.e.,  $\rho_{r_e}=\rho_{r_i}=\rho_{r}=0$ or $\rho_{r_e}=\rho_{r_i}=\rho_{r}=1$).
\end{itemize}

In the network represented in Fig.~\ref{fig:CVM} there are five hypocrisies.

There is a complete network in the external layer of CVM, and the two-layer network satisfies Properties~1-2 given above. Hypocrisy and cognitive dissonance can be reduced  by externalization or internalization. Within externalization, the hypocrisy will express an opinion publicly, and within internalization, the hypocrisy accepts his previously expressed opinion.

If we are interested in the number of hypocrisies, we can focus on  $\rho_{r}$  describing the stochastic matrix $Q_{CVM}$ in the CVM system, or we can describe the number of individuals having any of the state from set $S$ (the latter is considered in \cite{gastner2018consensus}). Suppose the external layer in the CVM system is a complete graph, then the number of individuals of each state is as follows:
\begin{equation}
  \begin{aligned}
    \label{eq:CVMnumber}
    N &= \#Rr+\#Rb+\#Br+\#Bb,\\
    r_e & = \#Rr+\#Rb,\\
    r_i & = \#Rr + \#Br,\\
    r & = \#Rr,\\
    \#Bb& = N-r_e-r_i+r,\\
    \#Rb&=r_e-r,\\
    \#Br&=r_i-r,
  \end{aligned}
\end{equation}
where $\#s$ is the number of individuals in state $s\in S$. 

We can represent the state of the system with $N$ individuals by triple $(\rho_{r_e}, \rho_{r_i}, \rho_{r})$. The transitions with positive probabilities from one state to other states for the described CVM system are given in Table~\ref{tab:CVMtransitions}. We use $Q$ to denote a transition matrix from here on.

\begin{sidewaystable}[htpb]
  \centering
  \caption{Transitions from state $(\rho_{r_e}, \rho_{r_i}, \rho_{r})$ in CVM}
  \label{tab:CVMtransitions}
  \begin{tabular}{p{2.8cm}p{6.2cm}p{4.5cm}p{4.5cm}}
    \toprule
    New state $(x,y,z)$& How can the new state be reached?& Probability&Transition rate matrix element $Q_{CVM}$ per unit time\\
    \midrule
    $(\rho_{r_e}+\frac{1}{N},\rho_{r_i}, \rho_{r})$ & $Bb\rightarrow Rb$, one individual with state $Bb$ copies a neighbor with external opinion $R$&$\frac{c(N-r_e-r_i+r)r_e}{N(N-1)+r_e+r_i-2r}$&$\frac{c(N-r_e-r_i+r)r_e}{N}$\\
    $(\rho_{r_e}+\frac{1}{N},\rho_{r_i}, \rho_{r}+\frac{1}{N})$ &$Br \rightarrow Rr$, one individual with state $Br$ copies a neighbor with external opinion $R$ or expresses his internal opinion&$\frac{c(r_i-r)r_e+e(r_i-r)}{N(N-1)+r_e+r_i-2r}$&$(r_i-r)(c\frac{r_e}{N}+e)$\\
    $(\rho_{r_e},\rho_{r_i}+\frac{1}{N}, \rho_{r}+\frac{1}{N})$ &$Rb \rightarrow Rr$, one individual with state $Rb$ accepts his external opinion & $\frac{i(r_e-r)}{N(N-1)+r_e+r_i-2r}$&$(r_e-r)i$\\
    $(\rho_{r_e}-\frac{1}{N},\rho_{r_i}, \rho_{r})$ &$ Rb \rightarrow Bb$, one individual with state $Rb$ copies a neighbor with external opinion $B$ or expresses his internal opinion &$\frac{c(r_e-r)(N-r_e)+e(r_e-r)}{N(N-1)+r_e+r_i-2r}$&$(r_e-r)(c \frac{N-r_e}{N}+e)$\\
    $(\rho_{r_e}-\frac{1}{N},\rho_{r_i}, \rho_{r}-\frac{1}{N})$ & $Rr \rightarrow Br$, one individual with state $Rr$ copies a neighbor with external opinion $B$ &$\frac{cr(N-r_e)}{N(N-1)+r_e+r_i-2r}$&$\frac{cr(N-r_e)}{N}$\\
    $(\rho_{r_e},\rho_{r_i}-\frac{1}{N}, \rho_{r})$ &$Br \rightarrow Bb$, one individual with state $Br$ accepts his external opinion &$\frac{i(r_i-r)}{N(N-1)+r_e+r_i-2r}$&$(r_i-r)i$\\
    $(\rho_{r_e},\rho_{r_i}, \rho_{r})$ &inverse to all above&one minus the sum above &negative sum of above elements\\
    \bottomrule
  \end{tabular}
\end{sidewaystable}

The updating procedure of CVM is similar to BVM, we can summarize the differences as follows:
\begin{itemize}
  \item Use a triple $(\rho_{r_e}, \rho_{r_i},\rho_{r})$ instead of $\rho$ to represent the state;
  \item As shown in Table~\ref{tab:CVMtransitions}, there are 6 possible state changes, we use the same scheme as in BVM, but with a different piecewise function to determine state changes;
\end{itemize}

Based on Table~\ref{tab:CVMtransitions}, we can describe the CVM system by expressing $\rho_{r}$, the proportion of hypocrisies in the network, given by equation~(\ref{eq:CVMstochastic}), where $Q_{CVM}[x,y]=\{P(x,y)\}$ represents the probability of transition from state $x$ (the previous value of $\rho_{r}$) to state $y$ (the next value of $\rho_{r}$). For $\rho_{r}\notin \{0,1\}$, we have equation~(\ref{eq:CVMstochastic}) and $P(0,0)=P(1,1)=1$:
\begin{equation}
  \begin{aligned}
    \label{eq:CVMstochastic}
    P\big(\rho_{r},\rho_{r}+\frac{1}{N}\big)  = &\, Q_{CVM}\Big[(\rho_{r_e}, \rho_{r_i}, \rho_{r}),\big(\rho_{r_e}+\frac{1}{N},\rho_{r_i}, \rho_{r}+\frac{1}{N}\big)\Big]\\
                                       &+ Q_{CVM}\Big[(\rho_{r_e}, \rho_{r_i}, \rho_{r}),\big(\rho_{r_e},\rho_{r_i}+\frac{1}{N}, \rho_{r}+\frac{1}{N}\big)\Big],\\
    P\big(\rho_{r},\rho_{r}-\frac{1}{N}\big) = &\, Q_{CVM}\Big[(\rho_{r_e}, \rho_{r_i}, \rho_{r}),\big(\rho_{r_e}-\frac{1}{N},\rho_{r_i}, \rho_{r}-\frac{1}{N}\big)\Big]\\
                                       P(\rho_{r},\rho_{r}) = &\, 1-P\big(\rho_{r},\rho_{r}+\frac{1}{N}\big)-P\big(\rho_{r},\rho_{r}-\frac{1}{N}\big).
  \end{aligned}
\end{equation}

\section{General Concealed Voter Model}{\label{sec3}}

\subsection{Motivation}

The General Concealed Voter Model (GCVM) idea is straightforward.
We assume that there exist connections in the internal layer.
It is reasonable for an individual to have his own close friends and share his real (internal) opinion with them.
Therefore, an individual and his friends may form either their own group or, a clique in the network. If the cliques are formed in the internal layer,
and individuals never share true opinions with others outside the clique except they are connected in the internal layer. First, we define the transition probabilities for a special case of internal network when it is represented by a complete graph, then we describe the dynamics of the network in the case of incomplete internal layer.

\subsection{Symmetric case: Complete internal network}

\begin{figure}[htpb]
  \centering
  \includegraphics[width=0.7\textwidth]{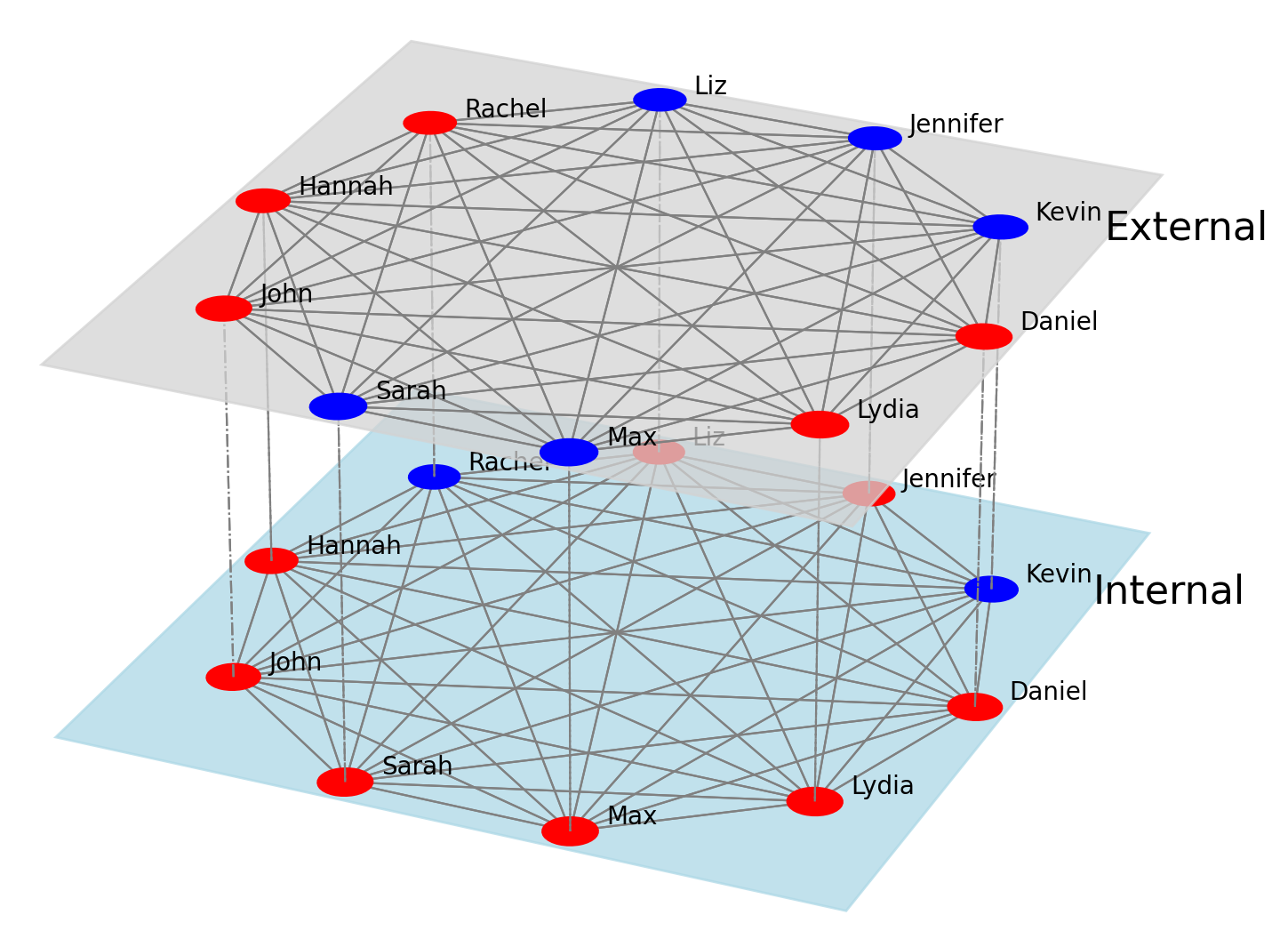}
  \caption{Representation of GCVM with 10 individuals: symmetric case}
  \label{fig:GCVM-1}
\end{figure}

First, we consider the simplest case of GCVM when both external and internal layers are complete networks. Figure~\ref{fig:GCVM-1} represents a symmetric case of GCVM with ten individuals.

The transition matrix is represented in  Table~\ref{tab:GCVMtransitions}.
There are eight cases of state transitions with nonzero probability while in the CVM system there are six cases, because  in the GCVM  system states $Bb$, $Rr$ are allowed to be changed to $Br$ and $Rb$ respectively caused by internal interactions.

\begin{sidewaystable}[htpb]
  \centering
  \caption{Transitions from state $(\rho_{r_e}, \rho_{r_i}, \rho_{r})$ in GCVM system with symmetric case}
  \label{tab:GCVMtransitions}
  \begin{tabular}{p{3.5cm}p{6.3cm}p{3.5cm}p{4.5cm}}
    \toprule
    New state $(x,y,z)$& How can the new state be reached?& Probability&Transition rate matrix element $Q_{GCVM}$ per unit time\\
    \midrule
    $(\rho_{r_e}+\frac{1}{N},\rho_{r_i}, \rho_{r})$ & $Bb\rightarrow Rb$, one individual with state $Bb$ copies a neighbor with external opinion $R$&$\frac{c(N-r_e-r_i+r)r_e}{2N(N-1)+r_e+r_i-2r}$&$\frac{c(N-r_e-r_i+r)r_e}{N}$\\
    $(\rho_{r_e}+\frac{1}{N},\rho_{r_i}, \rho_{r}+\frac{1}{N})$ &$Br \rightarrow Rr$, one individual with state $Br$ copies a neighbor with external opinion $R$ or expresses his internal opinion&$\frac{c(r_i-r)r_e+e(r_i-r)}{2N(N-1)+r_e+r_i-2r}$&$(r_i-r)(c\frac{r_e}{N}+e)$\\
    $(\rho_{r_e},\rho_{r_i}+\frac{1}{N}, \rho_{r}+\frac{1}{N})$ &$Rb \rightarrow Rr$, one individual with state $Rb$ accepts his external opinion or copies a neighbor with internal opinion $r$& $\frac{i(r_e-r)+c(r_e-r)r_i}{2N(N-1)+r_e+r_i-2r}$&$(r_e-r)(i+c \frac{r_i}{N})$\\
    $(\rho_{r_e},\rho_{r_i}+\frac{1}{N}, \rho_{r})$ &$Bb \rightarrow Br$, one individual with state $Bb$ copies a neighbor with internal opinion $r$& $\frac{c(N-r_e-r_i+r)r_i}{2N(N-1)+r_e+r_i-2r}$&$\frac{c(N-r_e-r_i+r)r_i}{N}$\\
    $(\rho_{r_e}-\frac{1}{N},\rho_{r_i}, \rho_{r})$ &$ Rb \rightarrow Bb$, one individual with state $Rb$ copies a neighbor with external opinion $B$ or expresses his internal opinion &$\frac{c(r_e-r)(N-r_e)+e(r_e-r)}{2N(N-1)+r_e+r_i-2r}$&$(r_e-r)(c \frac{N-r_e}{N}+e)$\\
    $(\rho_{r_e}-\frac{1}{N},\rho_{r_i}, \rho_{r}-\frac{1}{N})$ & $Rr \rightarrow Br$, one individual with state $Rr$ copies a neighbor with external opinion $B$ &$\frac{cr(N-r_e)}{2N(N-1)+r_e+r_i-2r}$&$\frac{cr(N-r_e)}{N}$\\
    $(\rho_{r_e},\rho_{r_i}-\frac{1}{N}, \rho_{r})$ & $Br \rightarrow Bb$, one individual with state $Br$ accepts his external opinion or copies a neighbor with internal opinion $b$&$\frac{i(r_i-r)+c(r_i-r)(N-r_i)}{2N(N-1)+r_e+r_i-2r}$&$(r_i-r)(i+c \frac{N-r_i}{N})$\\
    $(\rho_{r_e},\rho_{r_i}-\frac{1}{N}, \rho_{r}-\frac{1}{N})$ &$Rr \rightarrow Rb$, one individual with state $Rr$ copies a neighbor with internal opinion $b$&$\frac{cr(N-r_i)}{2N(N-1)+r_e+r_i-2r}$&$\frac{cr(N-r_i)}{N}$\\
    $(\rho_{r_e},\rho_{r_i}, \rho_{r})$ &inverse to all above&one minus the above sum &negative sum of above elements\\
    \bottomrule
  \end{tabular}
\end{sidewaystable}

Compared to CVM, new GCVM assumes internal interactions which influence the internal opinion transition process. In CVM, the changes of internal opinion can only happen through internalization (an individual accepts his external opinion), while in GCVM, internal opinion can be changed through internalization or interaction between internal nodes (or individuals' interaction in the internal layer). For example, nonhypocrisy individuals (individuals with states $Bb$ or $Rr$) can not change their states to $Br, Rb$ respectively in CVM system, but it is possible in GCVM. Additionally, internal interaction will prolong the consensus time and change the probability of each opinion winning.

Similarly, we can represent the stochastic matrix with respect to $\rho_r$.
For $\rho_{r}\notin \{0,1\}$, we have calculated  transition probabilities by~(\ref{eq:GCVMstochastic}) with $P(0,0)=P(1,1)=1$:
\begin{equation}
  \begin{aligned}
    \label{eq:GCVMstochastic}
    P\big(\rho_{r},\rho_{r}+\frac{1}{N}\big)  = & Q_{GCVM}\Big[(\rho_{r_e}, \rho_{r_i}, \rho_{r}),\big(\rho_{r_e}+\frac{1}{N},\rho_{r_i}, \rho_{r}+\frac{1}{N}\big)\Big]\\
                                       &+ Q_{GCVM}\Big[(\rho_{r_e}, \rho_{r_i}, \rho_{r}),\big(\rho_{r_e},\rho_{r_i}+\frac{1}{N}, \rho_{r}+\frac{1}{N}\big)\Big],\\
    P\big(\rho_{r},\rho_{r}-\frac{1}{N}\big) = &Q_{GCVM}\Big[(\rho_{r_e}, \rho_{r_i}, \rho_{r}),\big(\rho_{r_e}-\frac{1}{N},\rho_{r_i}, \rho_{r}-\frac{1}{N}\big)\Big]\\
                                       &+Q_{GCVM}\Big[(\rho_{r_e}, \rho_{r_i}, \rho_{r}),\big(\rho_{r_e},\rho_{r_i}-\frac{1}{N}, \rho_{r}-\frac{1}{N}\big)\Big],\\
                                       P(\rho_{r},\rho_{r}) = &1-P\big(\rho_{r},\rho_{r}+\frac{1}{N}\big)-P\big(\rho_{r},\rho_{r}-\frac{1}{N}\big).
  \end{aligned}
\end{equation}
We use the same measure as in \cite{gastner2018consensus, gastner2019impact} to represent the strength of the red opinion, that is
\begin{equation}
  \label{eq:m}
  m(\rho_{r_e},\rho_{r_i})=\frac{i\rho_{r_e}+e\rho_{r_i}}{e+i},
\end{equation}
where $m\in [0,1]$ is the weighted value that combines the proportion of red opinions in the internal and external layers, and  $m$ can be considered as an equivalent to $\rho$ in BVM.

\subsection{Asymmetric cases: Incomplete internal network}

In this section, we consider several asymmetric cases including internal star-coupled network, internal two star-coupled networks, and internal two-clique networks while the external layer is always the complete network. We assume that there are two groups which are equal in size both in the internal two star-coupled and two-clique networks.

\subsubsection{Internal layer: Star-coupled network}

\begin{figure}[htpb]
  \centering
  \includegraphics[width=0.7\textwidth]{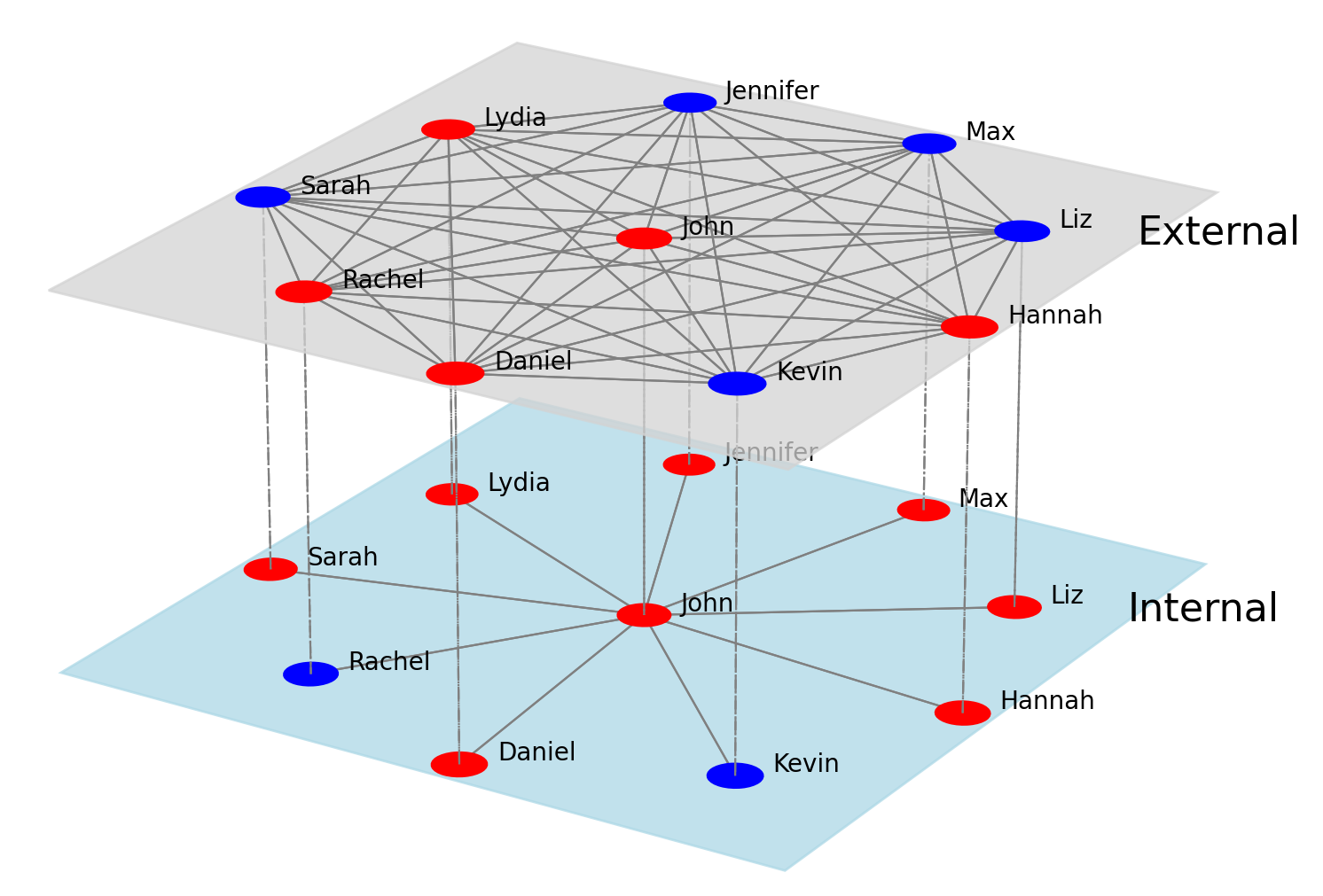}
  \caption{Representation of an internal star-coupled network}
  \label{fig:complete-star}
\end{figure}

In this case, we have a central node in the internal layer (star graph) as represented in Fig.~\ref{fig:complete-star}. In this case, the interaction in the internal layer exists only between the central node and any noncentral node. There is no interaction between noncentral nodes.

For the internal star-coupled network, there are 8 types of transitions with nonzero probabilities described in Table \ref{tab:one-star-transition}.

\begin{sidewaystable}
    \caption{Transitions from  state $(\rho_{r_e}, \rho_{r_i}, \rho_{r})$ in GCVM with internal star-coupled network}
    \label{tab:one-star-transition}
\begin{tabular}{p{3.5cm}p{9cm}p{5cm}}
    \toprule
    New state $(x,y,z)$& How can the new state be reached?& Transition rate matrix element $Q_{GCVM}$ per unit time\\
    \midrule
    $(\rho_{r_e}+\frac{1}{N},\rho_{r_i}, \rho_{r})$ & $Bb\rightarrow Rb$, one individual with state $Bb$ copies a neighbor with external opinion $R$&$\frac{c(N-r_e-r_i+r)r_e}{N}$\\
    $(\rho_{r_e}+\frac{1}{N},\rho_{r_i}, \rho_{r}+\frac{1}{N})$ & $Br \rightarrow Rr$, one individual with state $Br$ copies a neighbor with external opinion $R$ or expresses his internal opinion&$(r_i-r)(c\frac{r_e}{N}+e)$\\
    $(\rho_{r_e},\rho_{r_i}+\frac{1}{N}, \rho_{r}+\frac{1}{N})$ &$Rb \rightarrow Rr$, the authority in the internal layer has opinion $Rb$, he accepts his external opinion or copies a neighbor with internal opinion $r$; other individuals have the opinion $Bb$, one of them accepts his external opinion, or copies the authority iff the authority has internal opinion $r$ & $\frac{1}{N}(r_e-r)(i+c \frac{r_i}{N})+\frac{N-1}{N}(r_e-r)(i+c\frac{r_i}{N^2})$\\
    $(\rho_{r_e},\rho_{r_i}+\frac{1}{N}, \rho_{r})$ &$Bb \rightarrow Br$, the authority in the internal layer has opinion $Bb$, he copies a neighbor with internal opinion $r$; other individuals have the opinion $Bb$, one of them copies the authority iff the authority has internal opinion $r$ & $\frac{1}{N} (N-r_e-r_i+r) c\frac{r_i}{N}$+$\frac{N-1}{N}(r_e-r)(i+c\frac{r_i}{N^2})$\\
    $(\rho_{r_e}-\frac{1}{N},\rho_{r_i}, \rho_{r})$ & $ Rb \rightarrow Bb$, one individual with state $Rb$ copies a neighbor with external opinion $B$ or express his internal opinion &$(r_e-r)(c \frac{N-r_e}{N}+e)$\\
    $(\rho_{r_e}-\frac{1}{N},\rho_{r_i}, \rho_{r}-\frac{1}{N})$ & $Rr \rightarrow Br$, one individual with state $Rr$ copies a neighbor with external opinion $B$ &$\frac{cr(N-r_e)}{N}$\\
    $(\rho_{r_e},\rho_{r_i}-\frac{1}{N}, \rho_{r})$ & $Br \rightarrow Bb$, the authority in the internal layer has opinion $Br$, he accepts his external opinion or copies a neighbor with internal opinion $b$; other individuals have the opinion $Br$, one of them accepts his external opinion, or copies the authority iff the authority has internal opinion $b$&$\frac{1}{N}(r_i-r)(i+c\frac{N-r_i}{N})$+$\frac{N-1}{N}(r_i-r)(i+c\frac{N-r_i}{N^2})$\\
    $(\rho_{r_e},\rho_{r_i}-\frac{1}{N}, \rho_{r}-\frac{1}{N})$ &$Rr \rightarrow Rb$, the authority in the internal layer has opinion $Rr$, he copies a neighbor with internal opinion $b$; other individuals have the opinion $Rr$, one of them copies the authority iff the authority has internal opinion $b$ &$\frac{r}{N} c \frac{N-r_i}{N}$+$\frac{r(N-1)}{N} c \frac{N-r_i}{N^2}$\\
    $(\rho_{r_e},\rho_{r_i}, \rho_{r})$ & inverse of all above& negative sum of above elements\\
    \bottomrule
\end{tabular}
\end{sidewaystable}

If we compare the transitions listed in Tables~\ref{tab:GCVMtransitions} and \ref{tab:one-star-transition}, we can conclude that the changes of the internal network structure influence  only the transition rate of these transitions: $Rb\rightarrow Rr$, $Bb\rightarrow Br$, $Br\rightarrow Bb$ and $Rr\rightarrow Rb$. Therefore, in the following sections, where we consider other internal network structures, we will only show the transition rates for these four cases, all other transition rates ($Bb\rightarrow Rb$, $Br\rightarrow Rr$, $Rb\rightarrow Bb$ and $Rr\rightarrow Br$) remain the same as in Table \ref{tab:GCVMtransitions}.

\subsubsection{Internal layer: Two star-coupled network}

Assume that there is a two star-coupled network in the internal layer, and
we consider two possible cases:
\begin{enumerate}
  \item Two star-coupled networks share one common node (the total number of nodes is an odd number).
  \item Two star-coupled networks share two common nodes (the total number of nodes is an even number).
\end{enumerate}

\begin{figure}[htpb]
  \begin{subfigure}{.5\textwidth}
    \centering
    \includegraphics[width=1\linewidth]{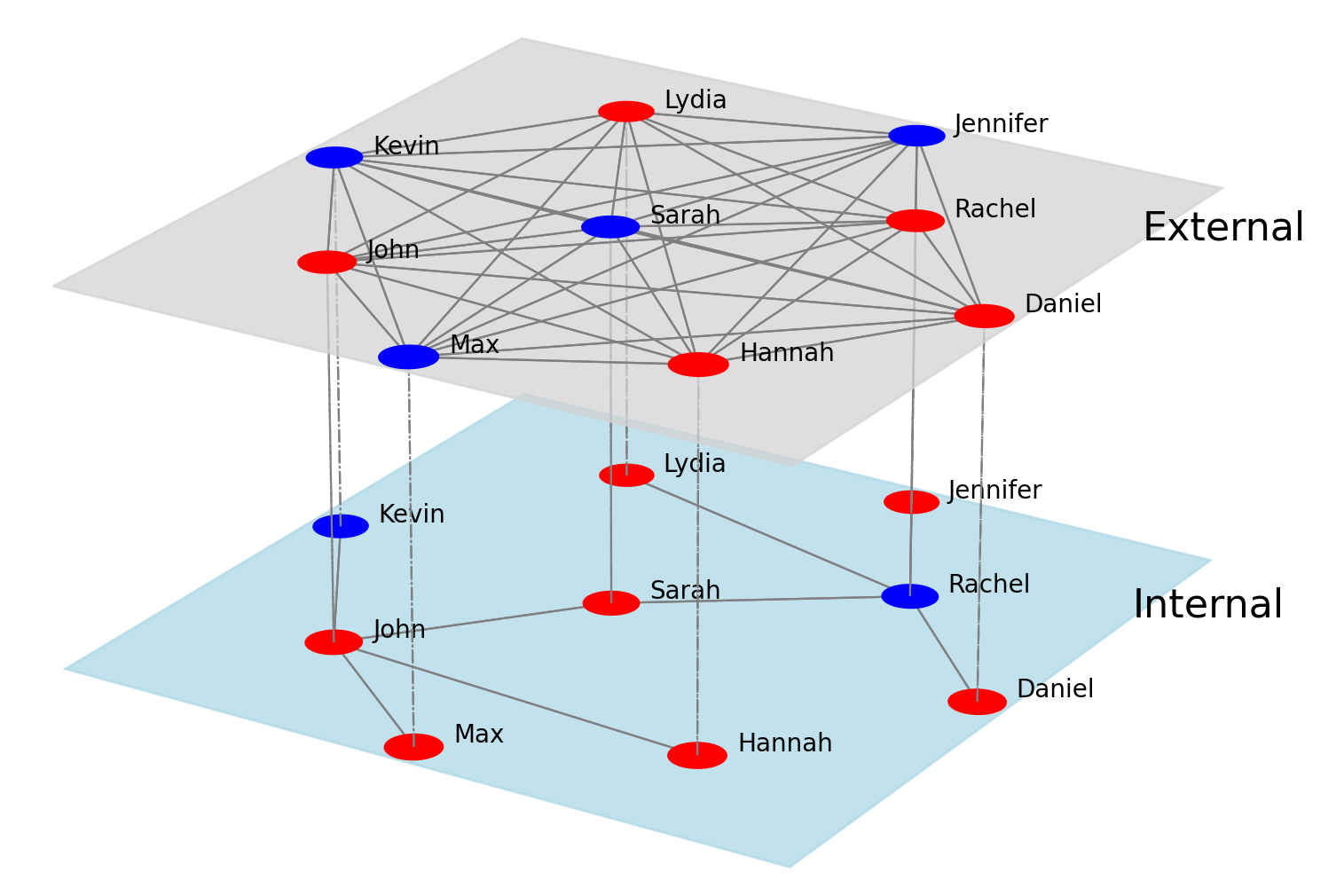}
    \caption{odd number of nodes}
    \label{fig:internal-two-star-odd}
  \end{subfigure}
  \begin{subfigure}{.5\textwidth}
    \centering
    \includegraphics[width=1\linewidth]{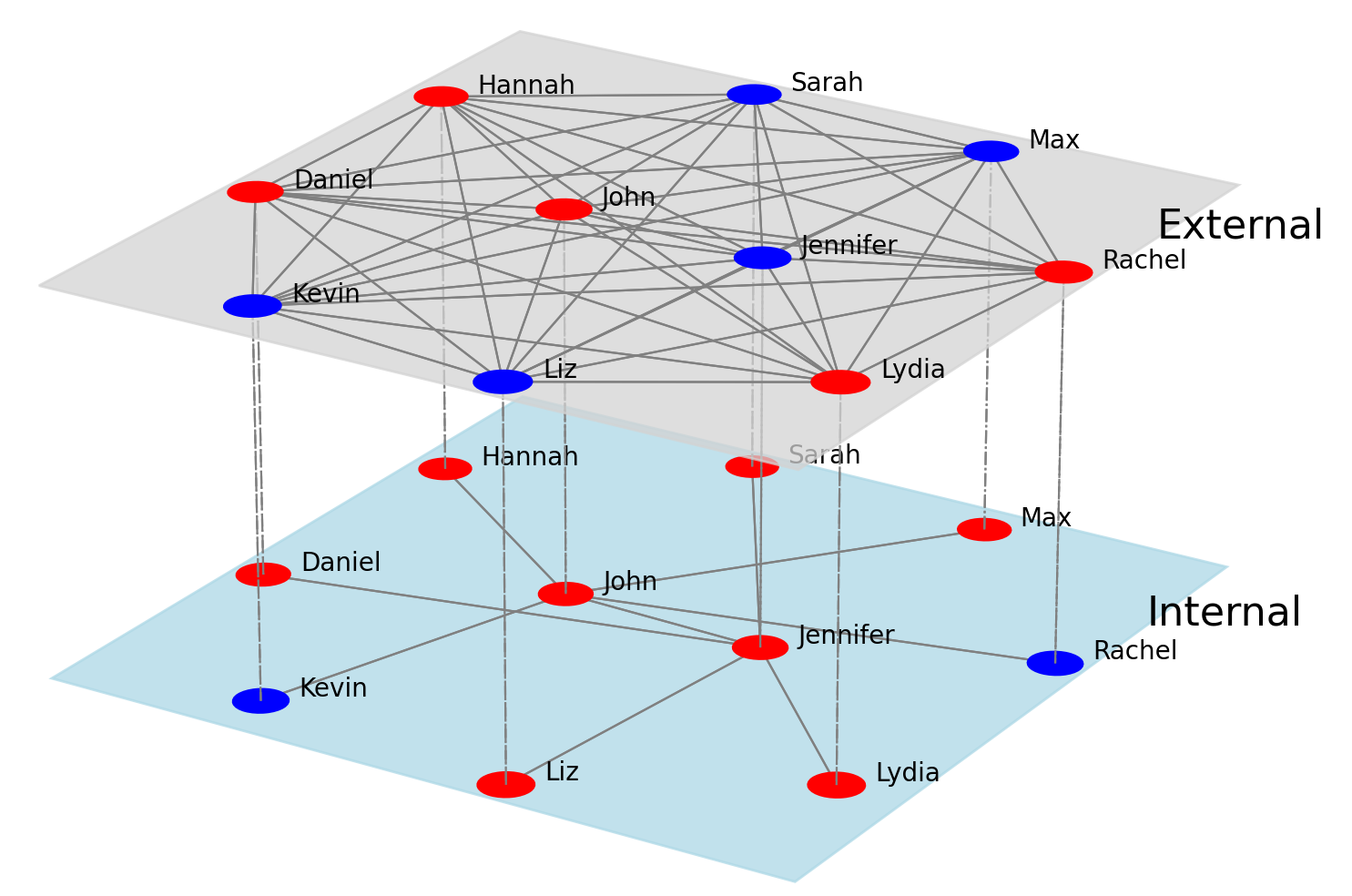}
    \caption{even number of nodes}
    \label{fig:internal-two-star-even}
  \end{subfigure}
  \caption{Representation of internal two star-coupled network}
  \label{fig:internal-two-star-networks}
\end{figure}

Fig.~\ref{fig:internal-two-star-odd} shows a simple example of the first case, each star subnetwork share a common node ``Sarah''. If we define a number of hubs (central nodes) by $a$, then for this case $a=2$. We also define $\rho_g$ as a proportion of nodes in one-star subnetwork of the nodes in the whole network, and $\rho_{g}=\frac{N+1}{2N}$. We characterize the transitions for the first case of two star-coupled internal network in Table~\ref{tab:two-star-odd-transition}.

\begin{sidewaystable}
    \caption{Transitions from state $(\rho_{r_e}, \rho_{r_i}, \rho_{r})$ in GCVM with two star-coupled internal networks (odd case)}
    \label{tab:two-star-odd-transition}
\begin{tabular}{p{3.5cm}p{9cm}p{5.5cm}}
    \toprule
    New state $(x,y,z)$& How can the new state be reached?& Transition rate matrix element $Q_{GCVM}$ per unit time\\
    \midrule
    $(\rho_{r_e},\rho_{r_i}+\frac{1}{N}, \rho_{r}+\frac{1}{N})$ &$Rb \rightarrow Rr$, the authority in the internal layer has opinion $Rb$, he accepts his external opinion or copies a neighbor with internal opinion $r$; other individuals have the opinion $Rb$, one of them accepts his external opinion, or copies the authority iff the authority has internal opinion $r$& $\frac{a}{N}(r_e-r)(i+c\frac{r_i}{N}\rho_{g})$+$\frac{N-a}{N}(r_e-r)i+\frac{N-a+1}{N}(r_e-r)c\frac{r_i}{N^2}$\\
    $(\rho_{r_e},\rho_{r_i}+\frac{1}{N}, \rho_{r})$ &$Bb \rightarrow Br$, the authority in the internal layer has opinion $Bb$, he copies a neighbor with internal opinion $r$; other individuals have the opinion $Bb$, one of them copies the authority iff the authority has internal opinion $r$&$\frac{a}{N} (N-r_e-r_i+r) c\frac{r_i}{N}\rho_g$+$\frac{N-a+1}{N}(N-r_e-r_i+r)c\frac{r_i}{N^2}$\\
    $(\rho_{r_e},\rho_{r_i}-\frac{1}{N}, \rho_{r})$ & $Br \rightarrow Bb$, the authority in the internal layer has opinion $Br$, he accepts his external opinion or copies a neighbor with internal opinion $b$; other individuals have the opinion $Br$, one of them accepts his external opinion, or copies the authority iff the authority has internal opinion $b$& $\frac{a}{N}(r_i-r)(i+c\frac{N-r_i}{N}\rho_g)$+$\frac{N-a}{N}(r_i-r)i$+$\frac{N-a+1}{N}(r_i-r)c\frac{N-r_i}{N^2}$\\
    $(\rho_{r_e},\rho_{r_i}-\frac{1}{N}, \rho_{r}-\frac{1}{N})$ &$Rr \rightarrow Rb$, the authority in the internal layer has opinion $Rr$, he copies a neighbor with internal opinion $b$; other individuals have the opinion $Rr$, one of them copies the authority iff the authority has internal opinion $b$ &$r\frac{a}{N} c \frac{N-r_i}{N}\rho_g$+$r\frac{(N-a+1)}{N} c \frac{N-r_i}{N^2}$\\
    \bottomrule
\end{tabular}
\end{sidewaystable}

Fig.~\ref{fig:internal-two-star-even} shows an example when in the internal layer, two star-coupled networks are connected by one edge through their hubs (central nodes)  \cite{belykh2005synchronization}. There are two common nodes ``John'' and ``Jennifer'' in each star subnetwork, and each star subnetwork has five nodes.
In this case, $\rho_g=\frac{N+2}{2N}$. We list the transitions in  Table~\ref{tab:two-star-even-transition}.

\begin{sidewaystable}
    \caption{Transitions from state $(\rho_{r_e}, \rho_{r_i}, \rho_{r})$ in GCVM with two star-coupled internal networks (even case)}
    \label{tab:two-star-even-transition}
\begin{tabular}{p{3.5cm}p{9cm}p{5.5cm}}
    \toprule
    New state $(x,y,z)$& How can the new state be reached?& Transition rate matrix element $Q_{GCVM}$ per unit time\\
    \midrule
    $(\rho_{r_e},\rho_{r_i}+\frac{1}{N}, \rho_{r}+\frac{1}{N})$ &$Rb \rightarrow Rr$,  the authority in the internal layer has opinion $Rb$, he accepts his external opinion or copies a neighbor with internal opinion $r$; other individuals have the opinion $Rb$, one of them accepts his external opinion, or copies the authority iff the authority has internal opinion $r$. &$\frac{a}{N}(r_e-r)(i+c\frac{r_i}{N}\rho_g)$+$\frac{N-a}{N}(r_e-r)(i+c\frac{r_i}{N^2})$\\
    $(\rho_{r_e},\rho_{r_i}+\frac{1}{N}, \rho_{r})$ &$Bb \rightarrow Br$, the authority in the internal layer has opinion $Bb$, he copies a neighbor with internal opinion $r$; other individuals have the opinion $Bb$, one of them copies the authority iff the authority has internal opinion $r$&$\frac{a}{N} (N-r_e-r_i+r) c\frac{r_i}{N}\rho_g$+$\frac{N-a}{N}(N-r_e-r_i+r)c\frac{r_i}{N^2}$\\
    $(\rho_{r_e},\rho_{r_i}-\frac{1}{N}, \rho_{r})$ & $Br \rightarrow Bb$, the authority in the internal layer has opinion $Br$, he accepts his external opinion or copies a neighbor with internal opinion $b$; other individuals have the opinion $Br$, one of them accepts his external opinion, or copies the authority iff the authority has internal opinion $b$& $\frac{a}{N}(r_i-r)(i+c\frac{N-r_i}{N}\rho_g)$+$\frac{N-a}{N}(r_i-r)(i+c\frac{N-r_i}{N^2})$\\
    $(\rho_{r_e},\rho_{r_i}-\frac{1}{N}, \rho_{r}-\frac{1}{N})$ &$Rr \rightarrow Rb$, the authority in the internal layer has opinion $Rr$, he copies a neighbor with internal opinion $b$; other individuals have the opinion $Rr$, one of them copies the authority iff the authority has internal opinion $b$.&$r\frac{a}{N} c \frac{N-r_i}{N}\rho_g$+$r\frac{(N-a)}{N} c \frac{N-r_i}{N^2}$\\
    \bottomrule
\end{tabular}
\end{sidewaystable}

\subsubsection{Internal layer: two-clique network}

Assume that there are two cliques in the internal layer, and there is one inter-clique link in this network. Each clique has the same number of individuals.
There are two possible cases:
\begin{enumerate}
  \item Two cliques share a common node (this common node has  degree $N-1$);
  \item Two cliques share two common nodes (these two common nodes have degree $\frac{2}{N}$).
\end{enumerate}

\begin{figure}[htpb]
  \begin{subfigure}{.5\textwidth}
    \centering
    \includegraphics[width=1\linewidth]{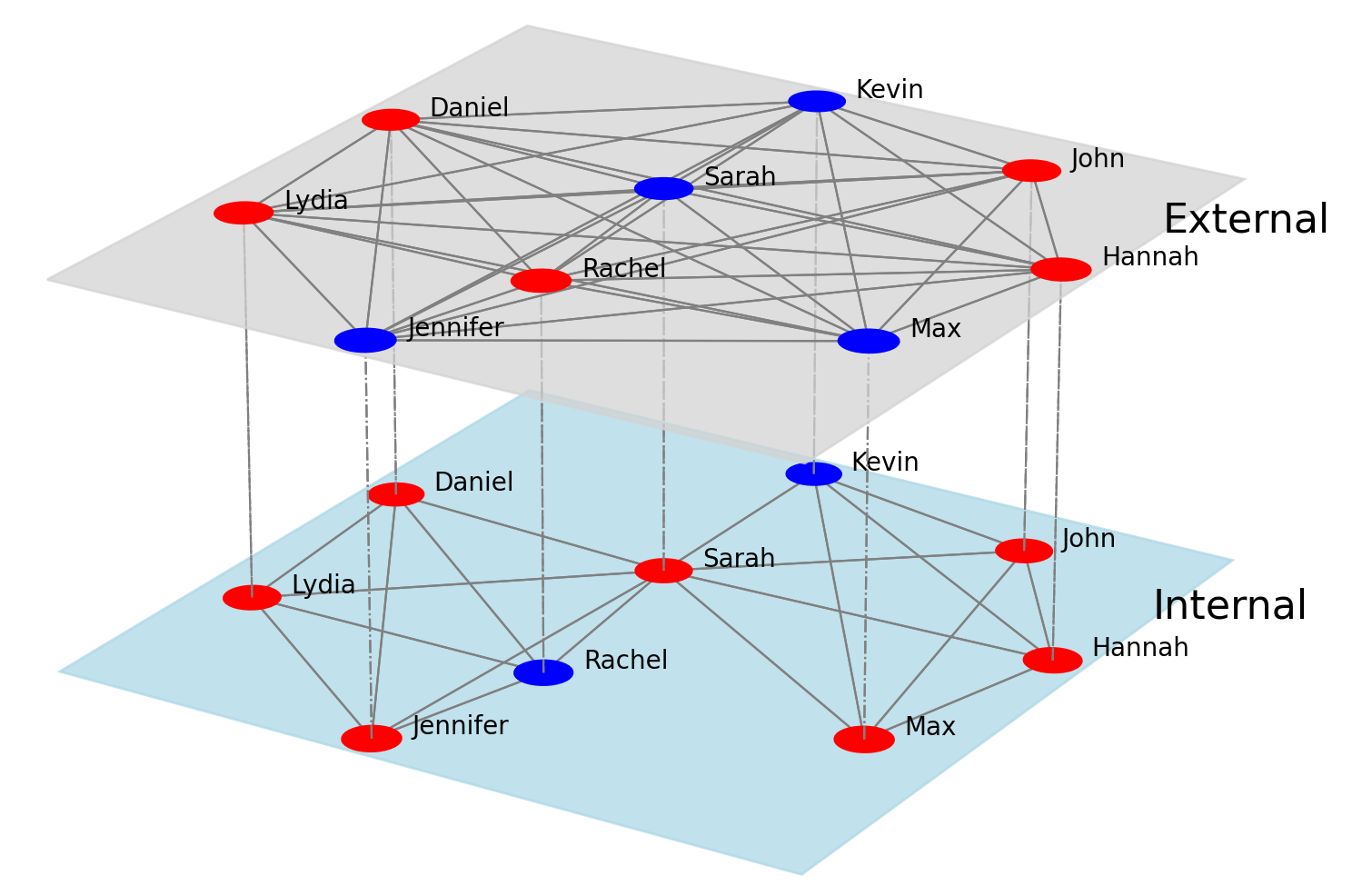}
    \caption{odd number of nodes}
    \label{fig:internal-two-clique-odd}
  \end{subfigure}
  \begin{subfigure}{.5\textwidth}
    \centering
    \includegraphics[width=1\linewidth]{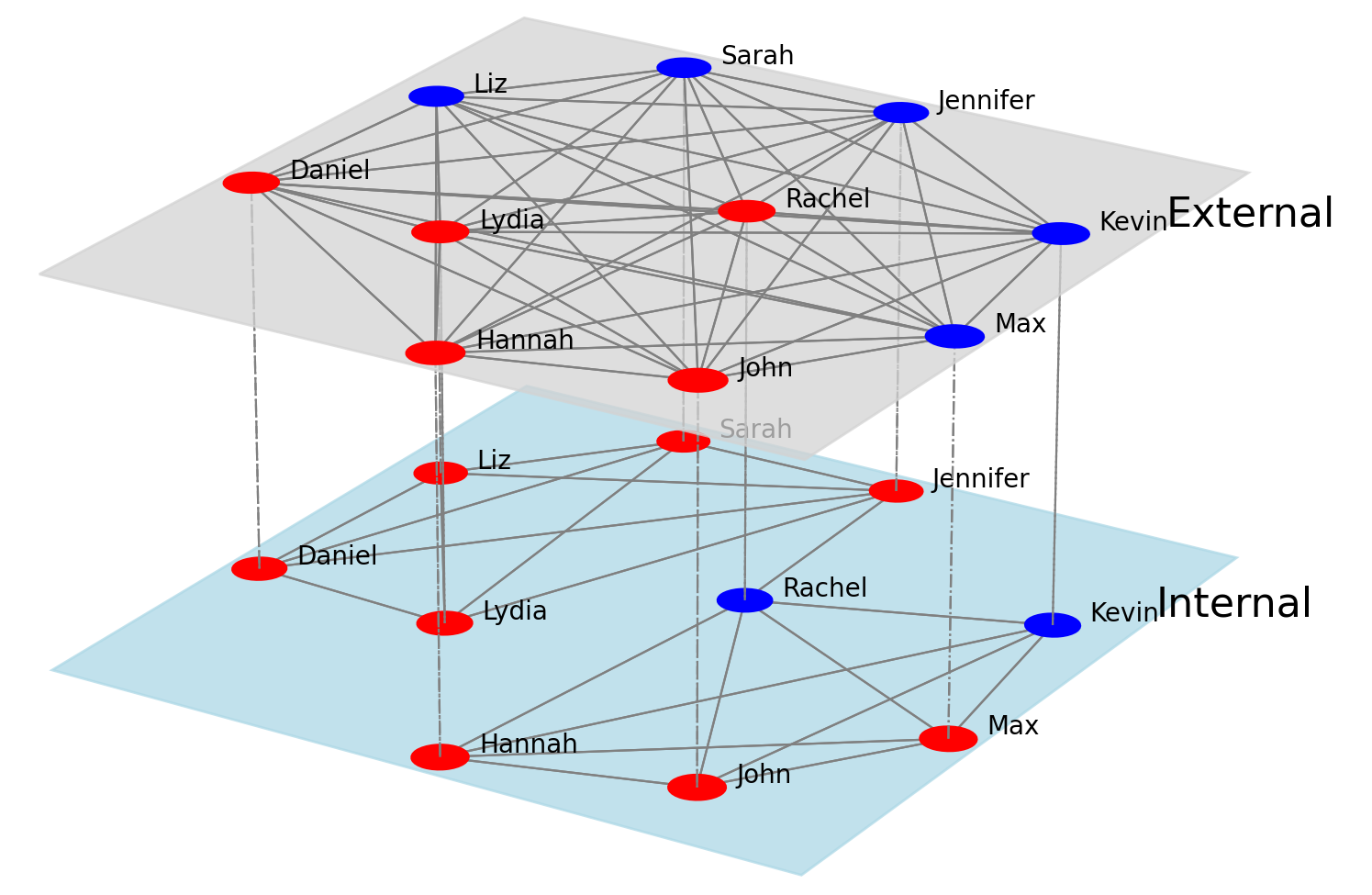}
    \caption{even number of nodes}
    \label{fig:internal-two-clique-even}
  \end{subfigure}
  \caption{Representation of internal two-clique network}
  \label{fig:internal-two-clique-networks}
\end{figure}

Fig.~\ref{fig:internal-two-clique-odd} shows an example of a two-clique network when two cliques share a common node ``Sarah''. The degree of node ``Sarah'' is $N-1$. Other nodes have the same degree $\frac{N-1}{2}$, and
$\rho_g=\frac{N+1}{2N}$ in this case. The transitions for this type of GCVM system are given in Table~\ref{tab:two-clique-odd-transition}.

Fig.~\ref{fig:internal-two-clique-even} shows a simple example of the second case of a two-clique internal network, when each clique has a special individual through whom two cliques are connected, i.e. there is a link connecting two special individuals, which is called an inter-clique link. In the second case, $\rho_g=\frac{N+2}{2N}$. The transition of this GCVM system are given in  Table~\ref{tab:two-clique-even-transition}.

\begin{sidewaystable}
    \caption{Transitions from the state $(\rho_{r_e}, \rho_{r_i}, \rho_{r})$ in GCVM with internal two clique-coupled networks (odd case)}
    \label{tab:two-clique-odd-transition}
\begin{tabular}{p{3.5cm}p{9cm}p{5.5cm}}
    \toprule
    New state $(x,y,z)$& How can the new state be reached?& Transition rate matrix element $Q_{GCVM}$ per unit time\\
    \midrule
    $(\rho_{r_e},\rho_{r_i}+\frac{1}{N}, \rho_{r}+\frac{1}{N})$ &$Rb \rightarrow Rr$,  the common node of the internal layer with opinion $Rb$ accepts his external opinion or copies a neighbor with internal opinion $r$; other individuals have the opinion $Rb$, one of them accepts his external opinion, or copies his neighbor's opinion. & $\frac{1}{N}(r_e-r)(i + c\frac{r_i}{N})$ + $\frac{N-1}{N}(r_e-r)(c\frac{r_i}{N}\rho_g+i)$\\
    $(\rho_{r_e},\rho_{r_i}+\frac{1}{N}, \rho_{r})$ &$Bb \rightarrow Br$, the common node of the internal layer has opinion $Bb$ copies a neighbor with internal opinion $r$; other individuals have the opinion $Bb$, one of them copies  his neighbor with internal opinion $r$.& $\frac{1}{N} (N-r_e-r_i+r) c\frac{r_i}{N}$ + $\frac{N-1}{N}(N-r_e-r_i+r)c\frac{r_i}{N}\rho_g$\\
    $(\rho_{r_e},\rho_{r_i}-\frac{1}{N}, \rho_{r})$ & $Br \rightarrow Bb$, the common node of the internal layer has opinion $Br$ accepts his external opinion or copies a neighbor with internal opinion $b$; other individuals have the opinion $Br$, one of them accepts his external opinion, or copies his neighbor with opinion $b$. & $\frac{1}{N}(r_i-r)(i+c\frac{N-r_i}{N})$+$\frac{N-1}{N}(r_i-r)(i+c\frac{N-r_i}{N}\rho_g)$\\
    $(\rho_{r_e},\rho_{r_i}-\frac{1}{N}, \rho_{r}-\frac{1}{N})$ &$Rr \rightarrow Rb$, the common node of the internal layer has opinion $Rr$ copies a neighbor with internal opinion $b$; other individuals have the opinion $Rr$, one of them copies his neighbor with internal opinion $b$. & $r\frac{1}{N} c \frac{N-r_i}{N}$+$r\frac{(N-1)}{N} c \frac{N-r_i}{N}\rho_g$\\
    \bottomrule
\end{tabular}
\end{sidewaystable}

\begin{sidewaystable}
    \caption{Transitions from state $(\rho_{r_e}, \rho_{r_i}, \rho_{r})$ in GCVM with internal two clique-coupled networks (even case)}
    \label{tab:two-clique-even-transition}
\begin{tabular}{p{3.5cm}p{9cm}p{5.5cm}}
    \toprule
    New state $(x,y,z)$& How is the new state reached?& Transition rate matrix element $Q_{GCVM}$ per unit time\\
    \midrule
    $(\rho_{r_e},\rho_{r_i}+\frac{1}{N}, \rho_{r}+\frac{1}{N})$ &$Rb \rightarrow Rr$, the common nodes of the internal layer have opinion $Rb$, one of them accepts its external opinion or copies a neighbor with internal opinion $r$; other individuals have the opinion $Rb$, one of them accepts his external opinion, or copies
 his neighbor's opinion.& $\frac{a}{N}(r_e-r)(i + c\frac{r_i}{N}\rho_g)$+$\frac{N-a}{N}(r_e-r)(c\frac{r_i}{aN}+i)$\\
    $(\rho_{r_e},\rho_{r_i}+\frac{1}{N}, \rho_{r})$ &$Bb \rightarrow Br$, common nodes of the internal layer have opinion $Bb$, one of them copies a neighbor with internal opinion $r$; other individuals have the opinion $Bb$, one of them copies  his
neighbor with internal opinion $r$.& $\frac{a}{N} (N-r_e-r_i+r) c\frac{r_i}{N}\rho_g$+$\frac{N-a}{N}(N-r_e-r_i+r)c\frac{r_i}{aN}$\\
    $(\rho_{r_e},\rho_{r_i}-\frac{1}{N}, \rho_{r})$ & $Br \rightarrow Bb$, common nodes of the internal layer have opinion $Br$, one of them accepts his external opinion or copies a neighbor with internal opinion $b$; other individuals have the opinion $Br$, one of them accepts his external opinion, or copies  his neighbor with opinion $b$.& $\frac{a}{N}(r_i-r)(i+c\frac{N-r_i}{N}\rho_g)$+$\frac{N-a}{N}(r_i-r)(i+c\frac{N-r_i}{aN})$\\
    $(\rho_{r_e},\rho_{r_i}-\frac{1}{N}, \rho_{r}-\frac{1}{N})$ &$Rr \rightarrow Rb$, common nodes of the internal layer have opinion $Rr$, one of them copies a neighbor with internal opinion $b$; other individuals have the opinion $Rr$, one of them copies his neighbor with internal opinion $b$. & $r\frac{a}{N} c \frac{N-r_i}{N}\rho_g$+$r\frac{(N-a)}{N} c \frac{N-r_i}{aN}$\\
    \bottomrule
\end{tabular}
\end{sidewaystable}

\subsection{Asymmetric case: Incomplete external network}

In this section, we assume that not only the internal layer but also the external layer can be incomplete graph.  These changes in the external layer will affect four 
transitions: $Bb\rightarrow Rb$, $Br\rightarrow Rr$,
$Rb\rightarrow Bb$, and $Rr\rightarrow Br$.

\subsubsection{External layer: Cycle network}

\begin{figure}[htpb]
  \begin{subfigure}{.3\textwidth}
    \centering
    \includegraphics[width=1\linewidth]{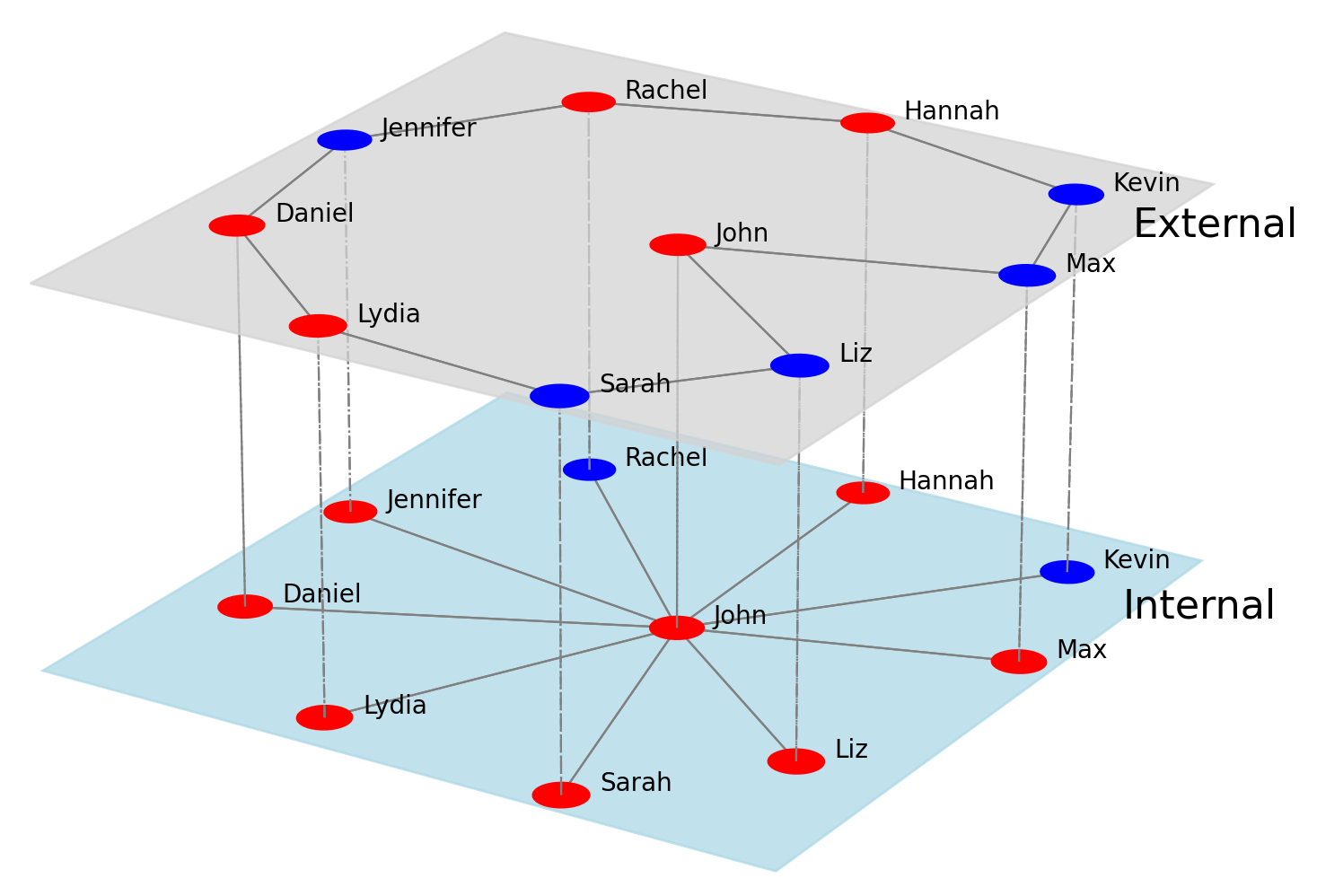}
    \caption{cycle-star}
    \label{fig:cycle-star}
  \end{subfigure}
  \begin{subfigure}{.3\textwidth}
    \centering
    \includegraphics[width=1\linewidth]{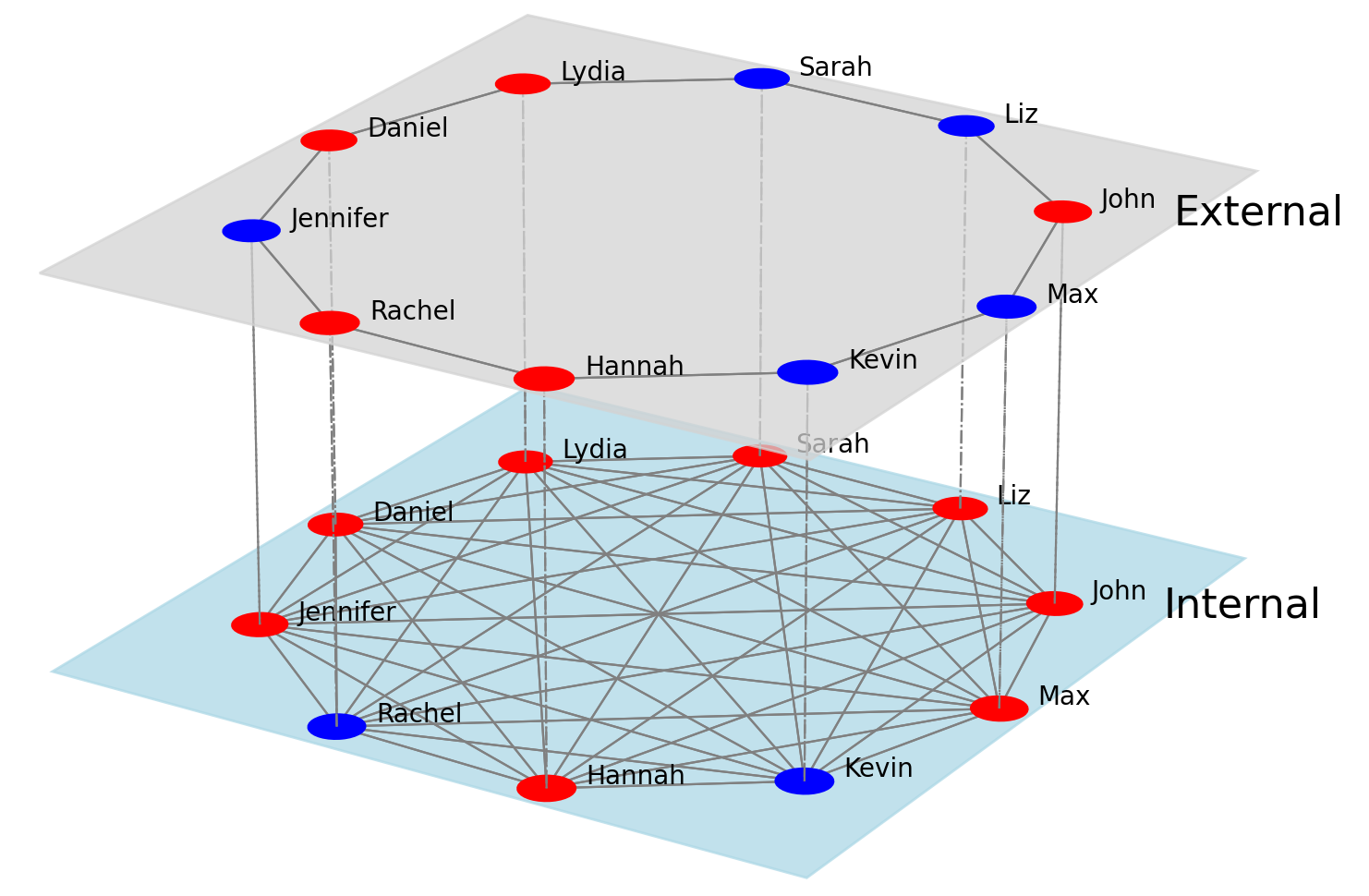}
    \caption{cycle-complete}
    \label{fig:cycle-star}
  \end{subfigure}
  \begin{subfigure}{.3\textwidth}
    \centering
    \includegraphics[width=1\linewidth]{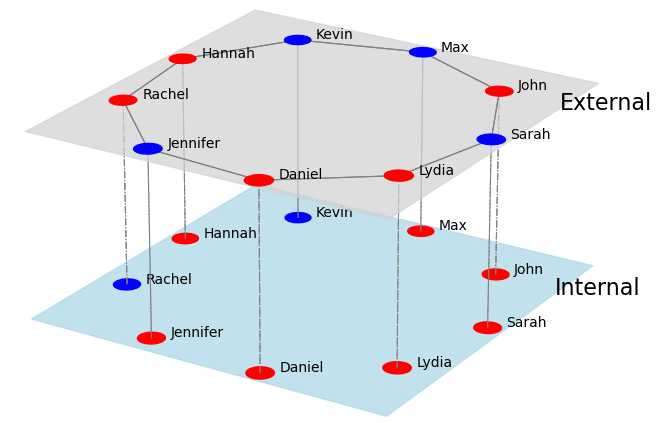}
    \caption{cycle-link free}
    \label{fig:cycle-linkFree}
  \end{subfigure}
  \caption{Representation of external cycle network}
  \label{fig:cycle-star-complete-linkfree}
\end{figure}

\begin{figure}[htpb]
  \begin{subfigure}{.5\textwidth}
    \centering
    \includegraphics[width=1\linewidth]{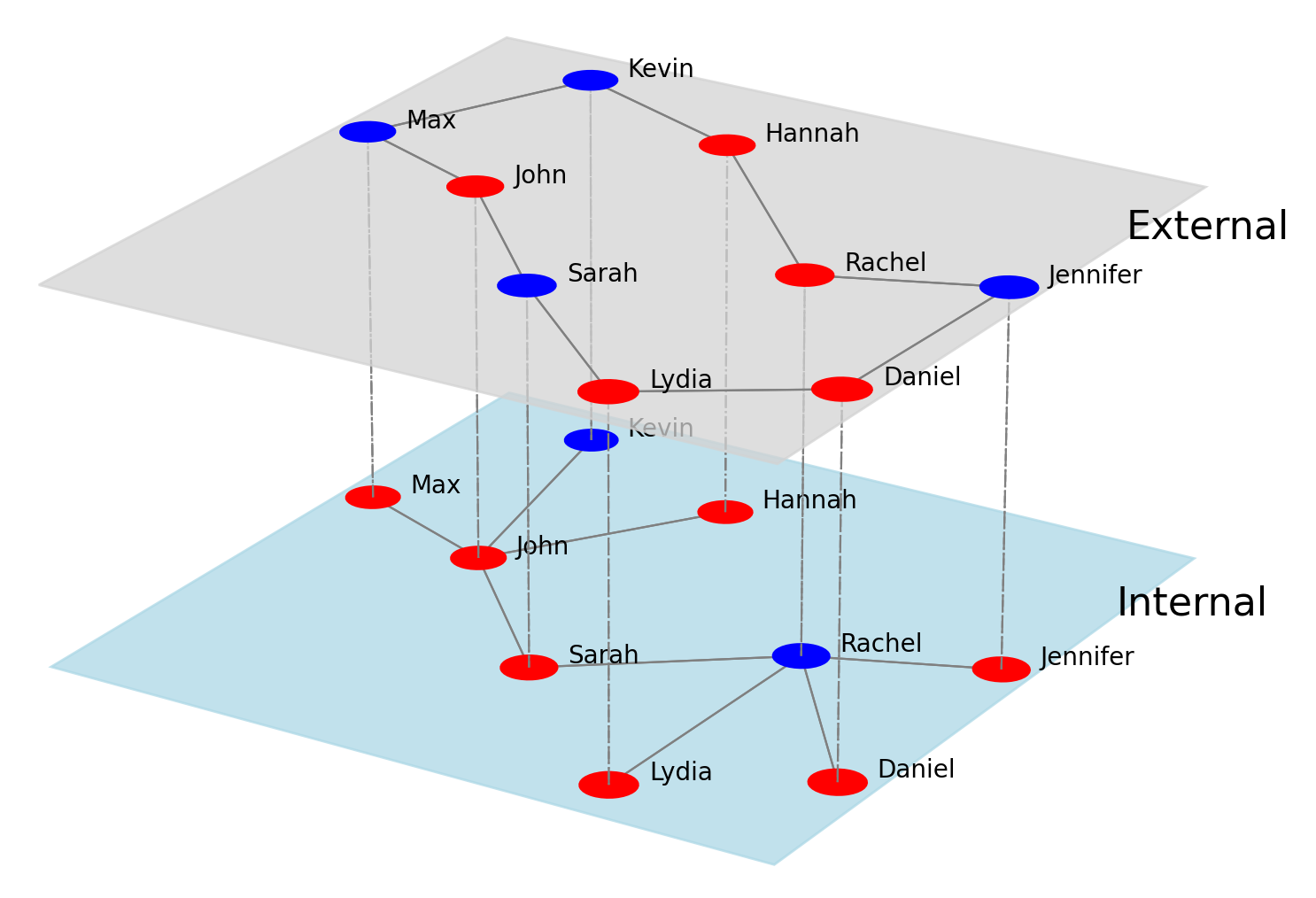}
    \caption{odd number of nodes}
    \label{fig:cycle-2star-odd}
  \end{subfigure}
  \begin{subfigure}{.5\textwidth}
    \centering
    \includegraphics[width=1\linewidth]{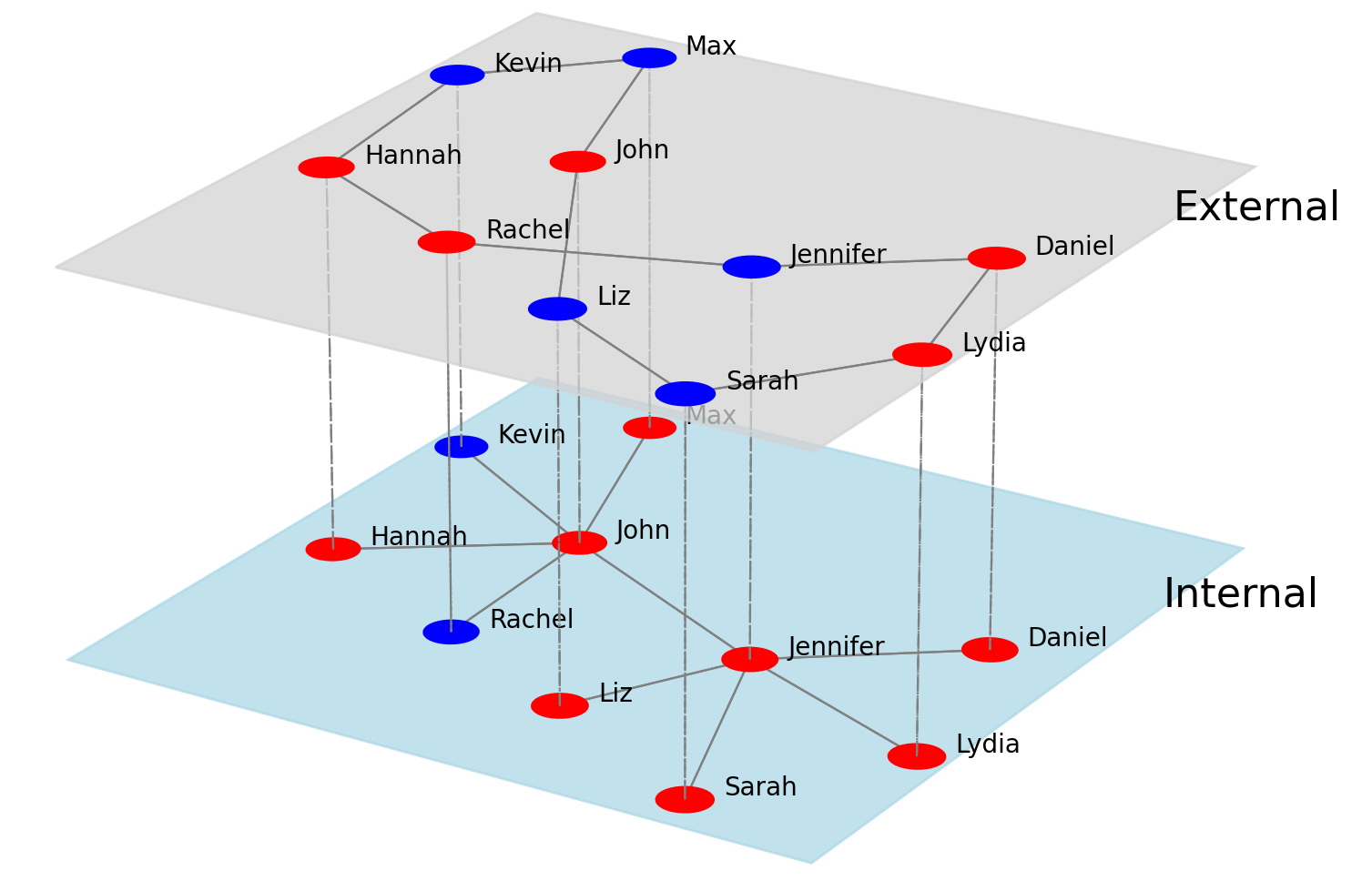}
    \caption{even number of nodes}
    \label{fig:cycle-2star-even}
  \end{subfigure}
  \caption{Representation of external cycle and internal two star-coupled network}
  \label{fig:cycle-2star}
\end{figure}

\begin{figure}[htpb]
  \begin{subfigure}{.5\textwidth}
    \centering
    \includegraphics[width=1\linewidth]{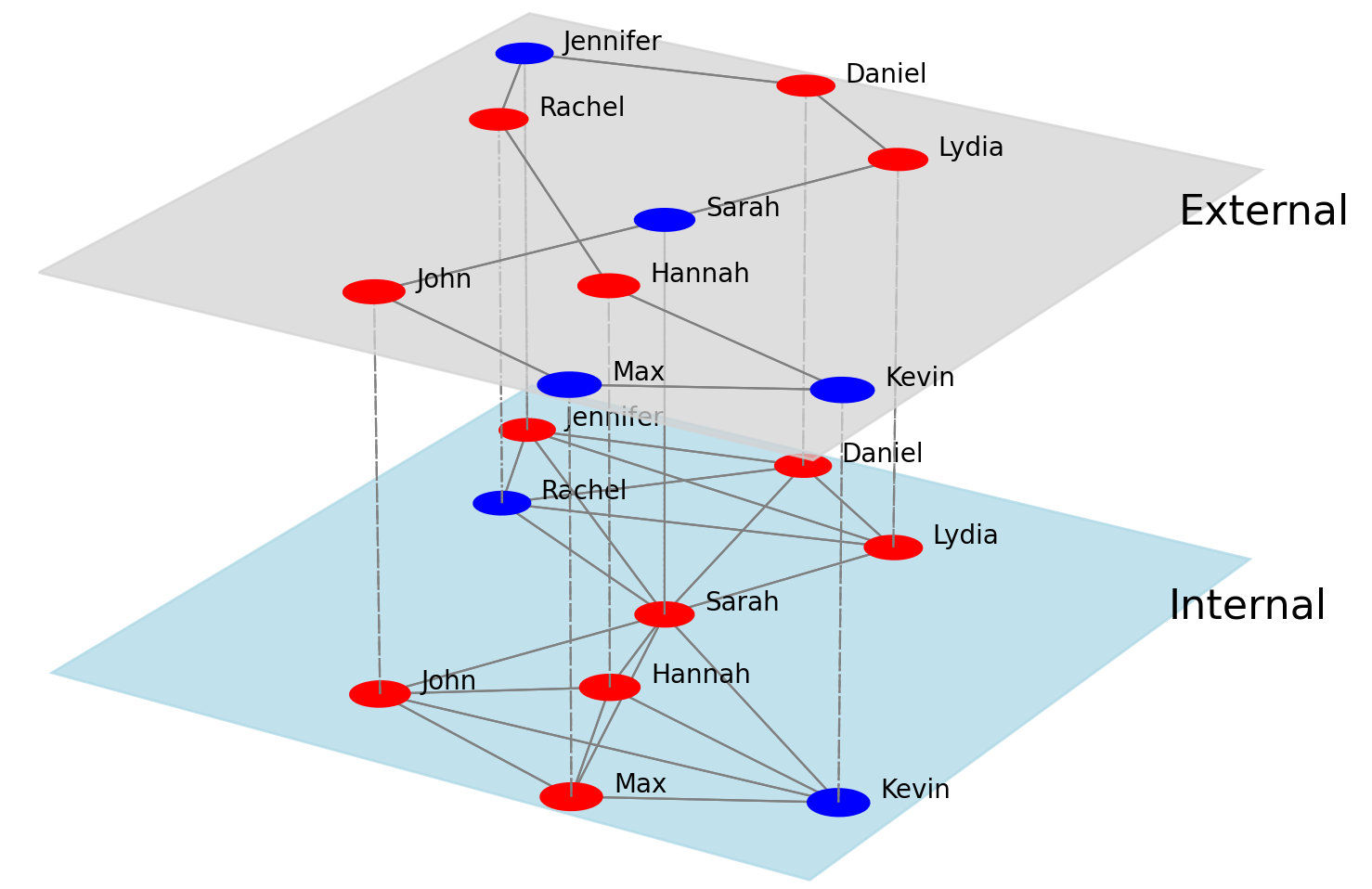}
    \caption{odd number of nodes}
    \label{fig:cycle-2clique-odd}
  \end{subfigure}
  \begin{subfigure}{.5\textwidth}
    \centering
    \includegraphics[width=1\linewidth]{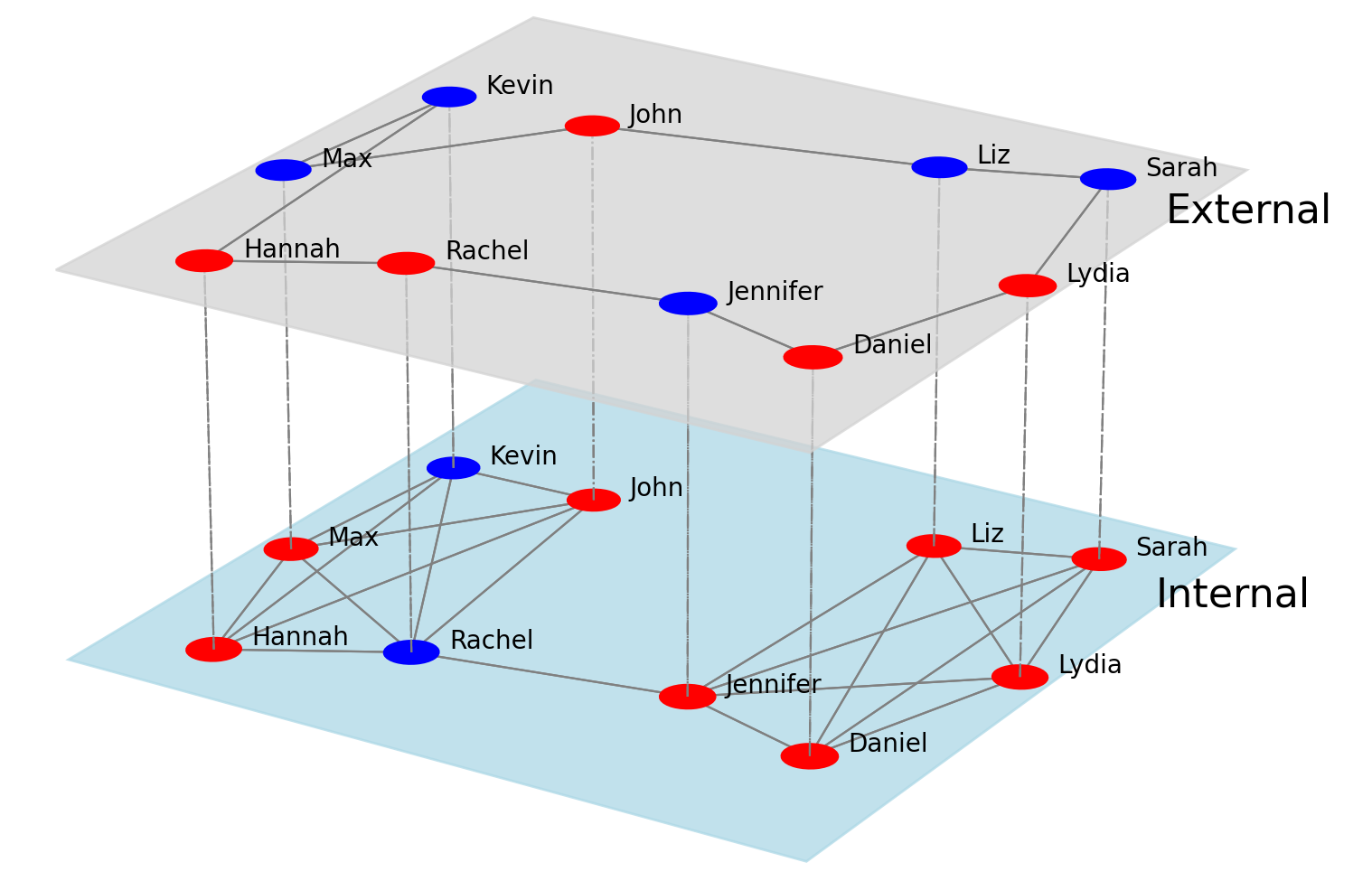}
    \caption{even number of nodes}
    \label{fig:cycle-2clique-even}
  \end{subfigure}
  \caption{Representation of external cycle and internal two-clique network}
  \label{fig:cycle-2clique}
\end{figure}

In this section, we assume that the external layer is given by a cycle. Then we represent how changes in the internal layer structure influence the transitions. We also consider the internal layers given by two-star and two-clique networks. The representations for such two-layer networks for GCVM are given in Figure~\ref{fig:cycle-star-complete-linkfree}, \ref{fig:cycle-2star}, \ref{fig:cycle-2clique}.

\begin{enumerate}
  \item $Bb\rightarrow Rb$: One individual with state $Bb$ copies a neighbor with external opinion $R$. The changing rate is $\frac{c(N-r_e-r_i+r)r_e}{N}\frac{3}{N}$;
  \item $Br\rightarrow Rr$: One individual with state $Br$ copies a neighbor with external opinion $R$, or expresses his internal opinion. The changing rate is $(r_i-r)(c \frac{r_e}{N}\frac{3}{N}+e)$;
  \item $Rb\rightarrow Bb$: One individual with state $Rb$ copies a neighbor with external opinion $B$, or expresses his internal opinion. The changing rate is $(r_e-r)(c \frac{N-r_e}{N} \frac{3}{N}+e)$;
  \item $Rr\rightarrow Br$: One individual with state $Rr$ copies a neighbor with external opinion $B$. The changing rate is $\frac{cr(N-r_e)}{N}\frac{3}{N}$.
\end{enumerate}

\section{Experiments and Results}{\label{sec4}}

We use the Monte Carlo method to simulate the process of opinion transition in two-layer network. The simulations are organised as follows:
\begin{enumerate}
  \item Set a random seed, the number of runs (denoted by $n$), and initialize the model by setting particular parameters;
  \item Verify whether the  consensus is reached for a given model (i.e. a unique opinion exists in the networks), and if so, stop the simulation, or else go to Step 3;
  \item Calculate the transition rate of each possible  state change under the current state;
  \item Determine the next state under the assumption that the time between two transitions is exponentially distributed with a current change rate;
  \item Go to Step 2.
\end{enumerate}

For simplicity, we refer to a consensus time and a winning rate as KPIs and examine them. The simulation is organized as follows:

\begin{enumerate}
  \item [1.] We run simulations varying parameters as shown below. Specifically, we focus on the strength of red opinion for each simulation, in this part we could see how the strength of red opinion influence the KPIs. Parameters used in the simulations are as follows:
    \begin{itemize}
      \item {\bf Set 1:} $\rho=0.75, N=400, r=300, c=1$ for BVM, and $m\approx 0.74, N=400,r_e=300, r_i=100, r=80, e=0.01, i=0.50, c=1$ for GCVM;
      \item {\bf Set 2:} $\rho=0.80, N=500, r=400, c=1$ for BVM, and $m\approx 0.79, N=500,r_e=400, r_i=100, r=80, e=0.01, i=0.80, c=1$ for GCVM;
    \end{itemize}
  \item [2.] We change the structure of internal layer and show how internal network structure influences the KPIs;
  \item [3.] We modify the structure of external layer and observe the impact of different external network structures on KPIs.
\end{enumerate}

The rest of this section demonstrates the most interesting results of our experiments.

\begin{figure}[htpb]
  \begin{subfigure}{.5\textwidth}
    \centering
    \includegraphics[width=1\linewidth]{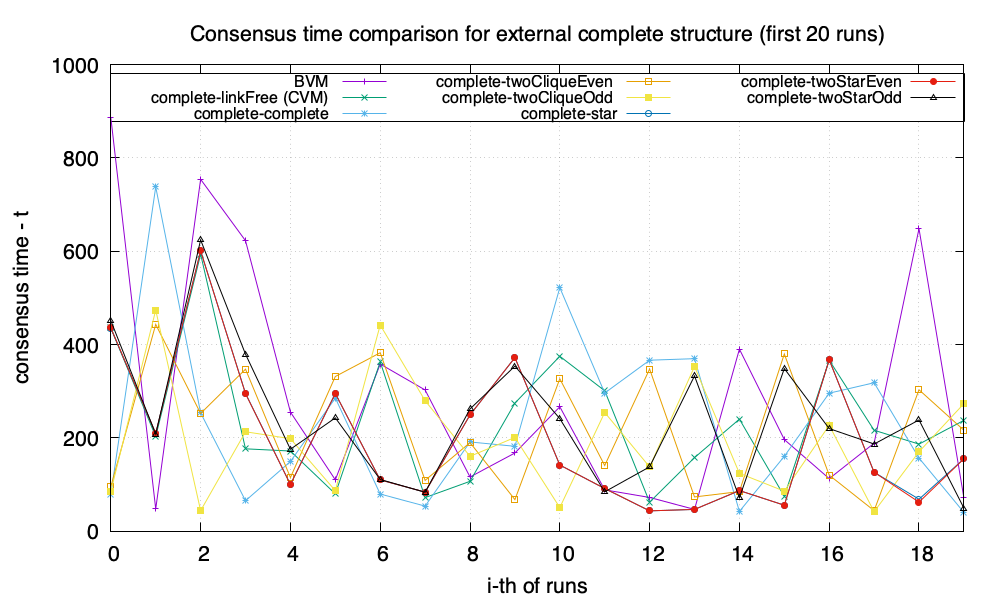}
    \caption{Parameter set 1}
    \label{fig:timeComparison_ecomplete_p1}
  \end{subfigure}
  \begin{subfigure}{.5\textwidth}
    \centering
    \includegraphics[width=1\linewidth]{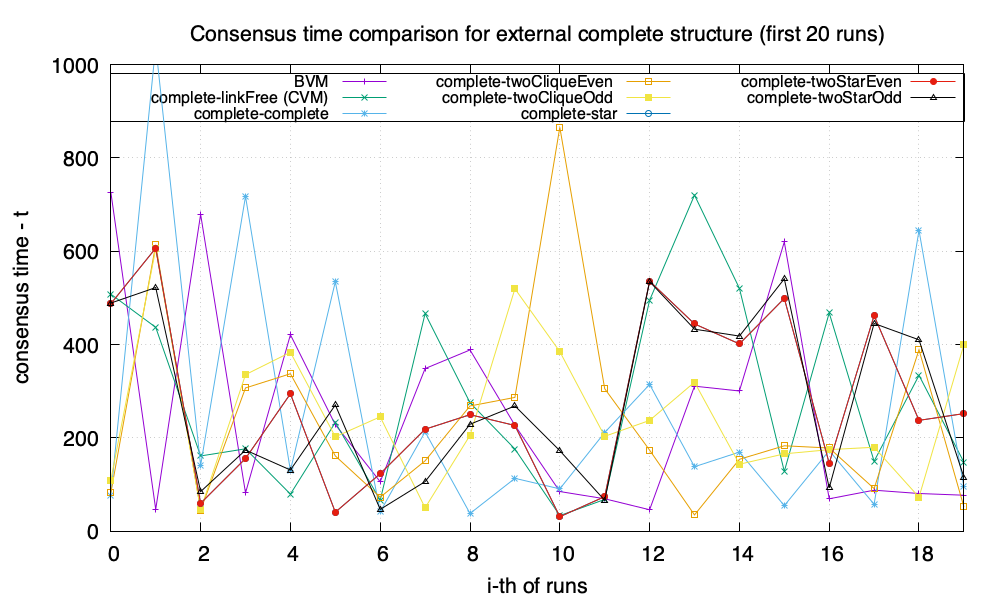}
    \caption{Parameter set 2}
    \label{fig:timeComparison_ecomplete_p2}
  \end{subfigure}
  \caption{Consensus time for 20 runs, comparison between BVM, CVM and GCVM with external complete network structure}
  \label{fig:timeComparison_ecomplete}
\end{figure}

Fig.~\ref{fig:timeComparison_ecomplete} shows the consensus time for the first 20 simulations. We can see that the consensus time of BVM, CVM and GCVM with external complete network structure is in the same order of magnitude.

\begin{figure}[htpb]
  \begin{subfigure}{.5\textwidth}
    \centering
    \includegraphics[width=1\linewidth]{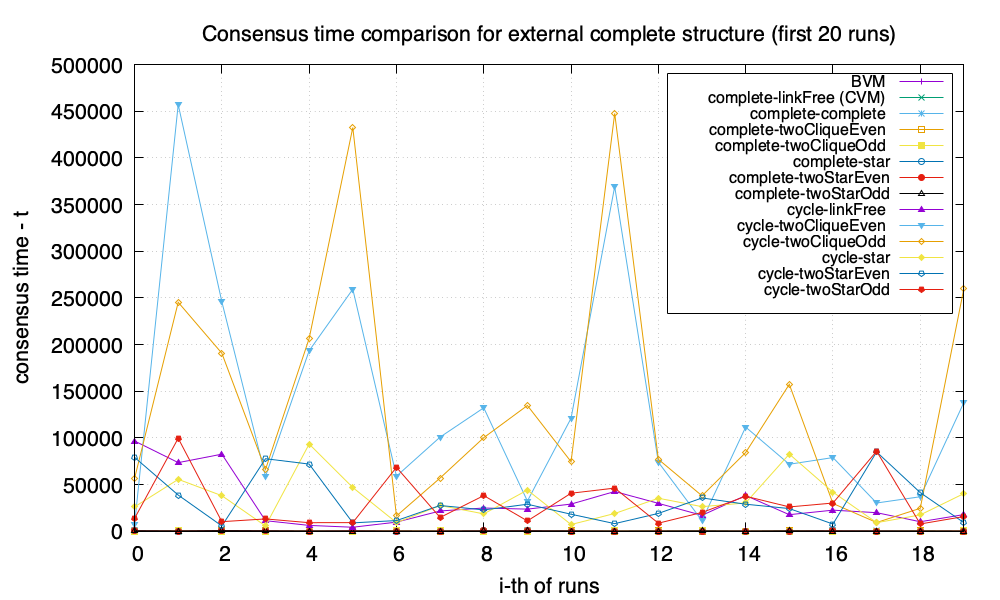}
    \caption{Parameter set 1}
    \label{fig:timeComparison_all_p1}
  \end{subfigure}
  \begin{subfigure}{.5\textwidth}
    \centering
    \includegraphics[width=1\linewidth]{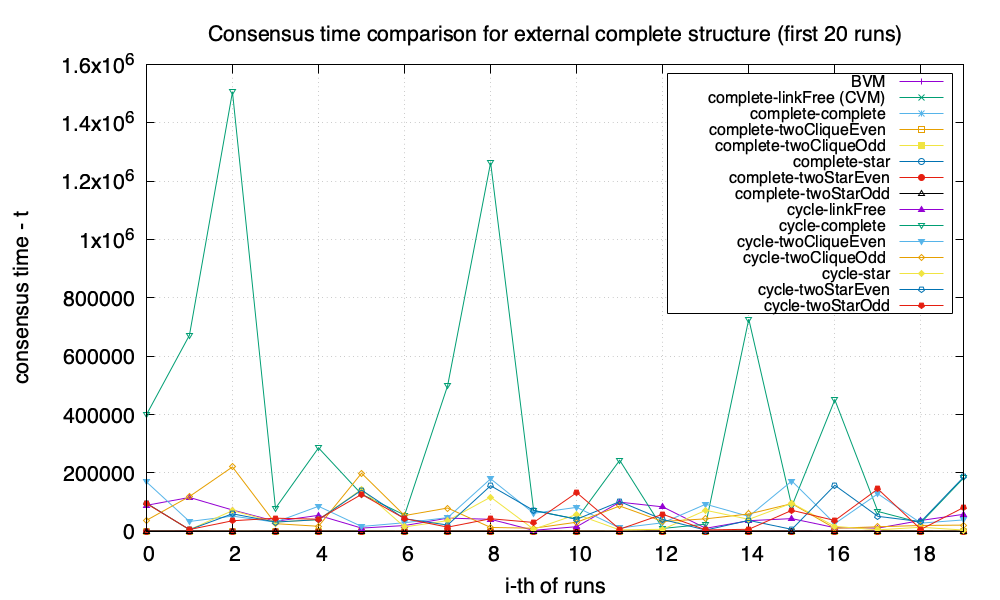}
    \caption{Parameter set 2}
    \label{fig:timeComparison_all_p2}
  \end{subfigure}
  \caption{Consensus time for 20 runs, comparison between all models}
  \label{fig:timeComparison_all}
\end{figure}

We can consider GCVM with incomplete external structure and compare the results with the previously described models. The results are summarized in Fig.~\ref{fig:timeComparison_all}. We highlight  the surprising result: in the external layer the cycle structure is assumed to be much simpler than the complete structure, but it greatly prolongs the consensus time. One can observe the difference between the left and right subfigures --- there is no ``cycle-complete'' model in the left subfigure. We do not put this model because in our simulations, there is no consensus on the model under parameter set 1.
Even under parameter set 2, the consensus time for a structure ``cycle-complete'' is much longer than for other models.

\begin{figure}[htpb]
  \begin{subfigure}{.5\textwidth}
    \centering
    \includegraphics[width=1\linewidth]{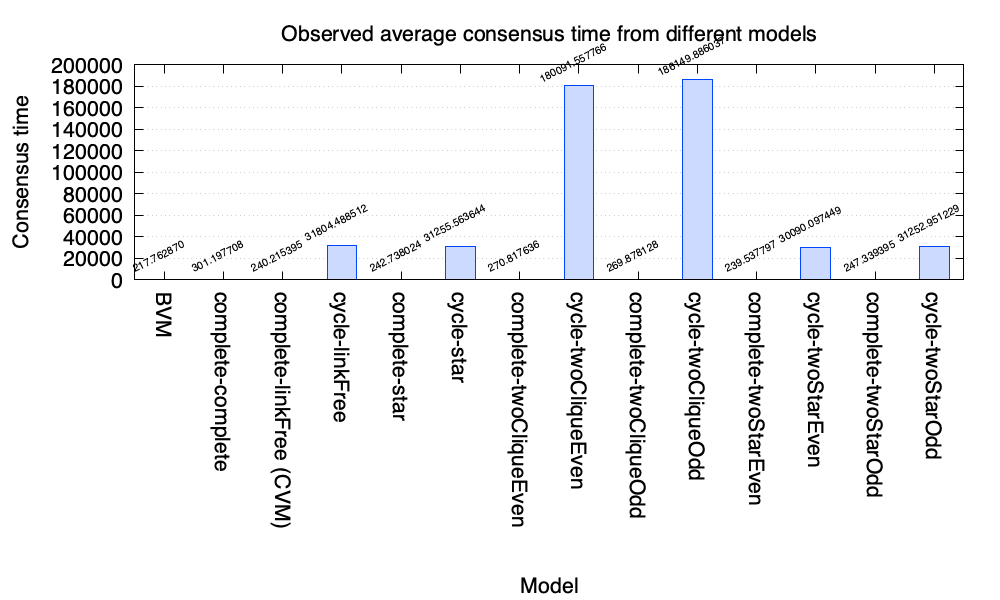}
    \caption{Parameter set 1}
    \label{fig:average_consensus_time_p1}
  \end{subfigure}
  \begin{subfigure}{.5\textwidth}
    \centering
    \includegraphics[width=1\linewidth]{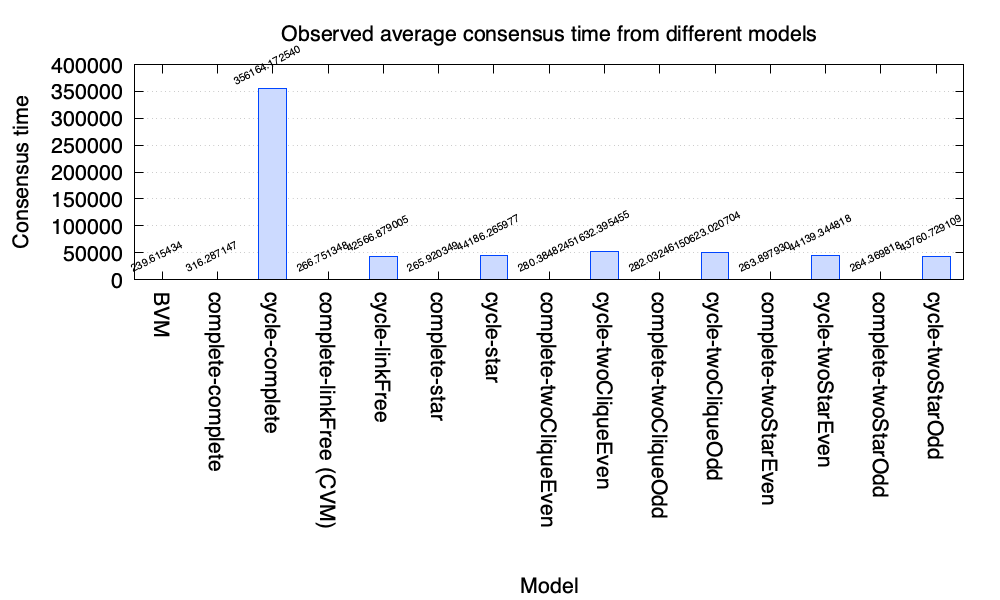}
    \caption{Parameter set 2}
    \label{fig:average_consensus_time_p2}
  \end{subfigure}
  \caption{Observed average consensus time for all models}
  \label{fig:average_consensus_time}
\end{figure}

Fig.~\ref{fig:average_consensus_time} represents the observed average consensus time for all models. It is not too hard to make the same conclusion as the one based on Fig.~\ref{fig:timeComparison_all} that the external layer cycle structure prolongs the consensus time significantly.

If we compare the observed average consensus time between BVM, CVM and GCVM with external complete structure, we can draw the following conclusions:
\begin{enumerate}
  \item Multi-layer network structure of the society prolongs the consensus time in comparison with  BVM, in which there is a unique communication layer;
  \item Not all internal structures prolong the consensus time in CVM.
\end{enumerate}

\begin{figure}[htpb]
  \begin{subfigure}{.5\textwidth}
    \centering
    \includegraphics[width=1\linewidth]{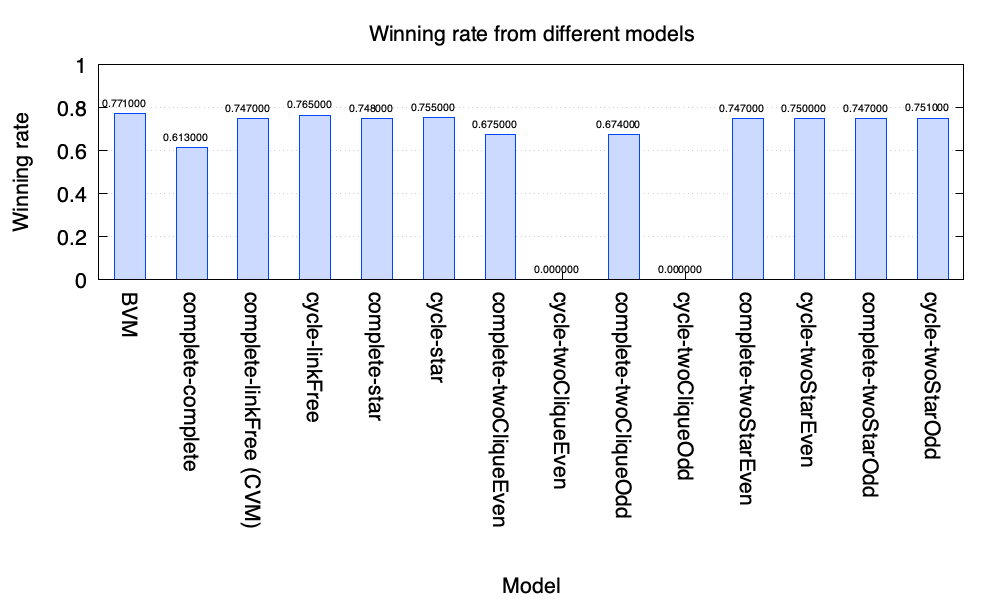}
    \caption{Parameter set 1}
    \label{fig:winning_rate_p1}
  \end{subfigure}
  \begin{subfigure}{.5\textwidth}
    \centering
    \includegraphics[width=1\linewidth]{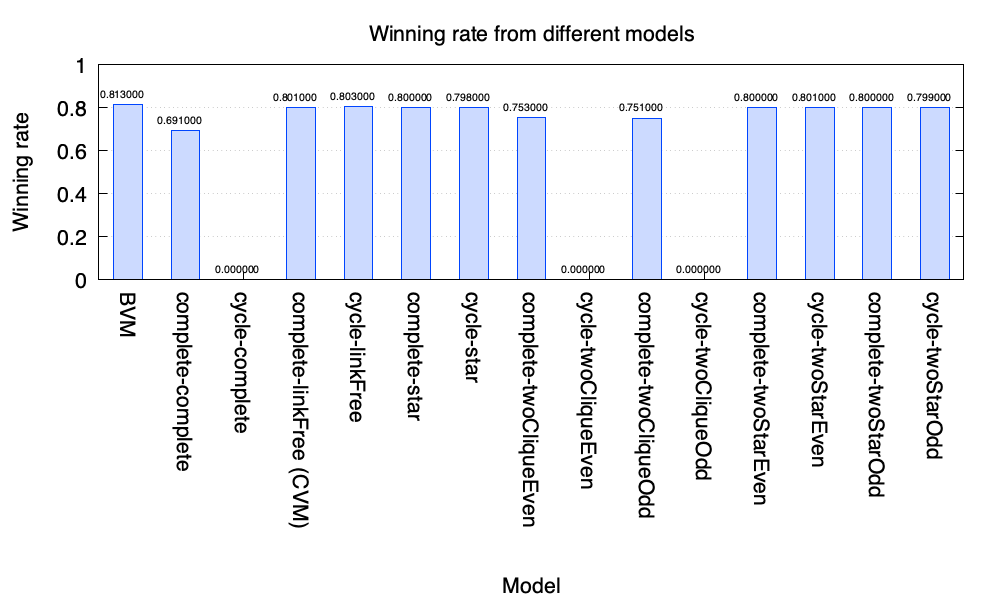}
    \caption{Parameter set 2}
    \label{fig:winning_rate_p2}
  \end{subfigure}
  \caption{Winning rate of red opinion for all models}
  \label{fig:winning_rate}
\end{figure}

We present Fig.~\ref{fig:winning_rate} based on 1000 simulations of each model. As the graph shows, BVM has the highest winning rate for both parameter sets. The case ``cycle-complete'' (external layer is cycle, internal is complete graph) is interesting because for parameter set 1 the consensus is not reached in our numerical simulations, therefore this structure is omitted in Fig. \ref{fig:average_consensus_time_p1}. In the case ``cycle-complete'' with the  parameter set 2, the consensus is reached the time to consensus is extremely high in comparison with all other structures of the layers. We also make an interesting observation: for the internal layer of a complex structure (complete, two-clique), the external cycle structure reduces the winning rate of the red opinion, which is close to 0, although the strength of the initial red opinion given by our parameters sets is very high. We can explain this as follows: when the external layer network structure is cyclic, the lower the average degree of nodes, the higher the consensus time, and the lower the winning rate. If we compare  Fig.~\ref{fig:winning_rate_p1} with Fig.~\ref{fig:winning_rate_p2}, we can observe the higher strength of the red opinion and its higher winning rate. Therefore, our conclusion seems to be reasonable.

\section{Conclusions}

This paper proposes a general concealed voter model (GCVM) which is an extension of CVM. Numerical simulations for GCVM with different internal and external structures, and various parameters of the model are made. The main formulated conclusions based on the numerical simulations including some counter-intuitive ones are as follows:
\begin{enumerate}
  \item For some simulations, a simple external layer network structure like cycle creates a problem for reaching a consensus;
  \item If individuals in the social network are not good at expressing their opinions publicly (have a low value of parameter $e$), internal interaction does not have a great influence on consensus (including the winning rate and consensus time).
\end{enumerate}
In the future, one can try to obtain the theoretical results on the consensus time and winning rate for each model depending on the internal and external layer network structures. We also find it interesting to examine more complex network structures which are the basis of individuals' interactions.

\section*{Acknowledgments}

The work of the second author was supported by National Natural Science Foundation of China (No. 72171126).

\section*{Declarations}

The authors have no relevant financial or non-financial interests to disclose. The authors equally contributed to this work.

\bibliographystyle{unsrtnat}
\bibliography{references}  






\end{document}